\documentstyle [12pt]{report}
\begin{document}
  \baselineskip=20pt

\begin{titlepage}

\phantom{boo}
\vskip 2.0cm
\centerline{\huge Field Theoretical Methods in Cosmology}

\vskip 1.0cm
\centerline{\large by}
\vskip 1.0cm
\centerline{\large Anupam Singh}

\vskip 7.0cm
\centerline{A thesis submitted in conformity with the requirements}
\centerline{for the degree of Doctor of Philosophy at the }
\centerline{Physics Department, Carnegie Mellon University }
\centerline{ 1995}

\vskip 2.0cm

\end{titlepage}


\begin{center}
{\bf Abstract}
\end{center}
\begin{quotation}
To optimally utilize all the
exciting cosmological data coming in we need
to sharpen also the theoretical tools available to cosmologists.
One such indispensible tool to understand hot big bang cosmology is
finite temperature field theory.
We review
and summarise the efforts made by us to use finite temperature field theory
to address issues of current interest to cosmologists.
An introduction to both the real time and the
imaginary time formalisms is provided.

The imaginary time formalism is illustrated by applying it
to understand the interesting possibilty of Late Time Phase Transitions.
Recent observations of the space distribution of quasars indicate
a very notable
peak in space density at a redshift of $2$ to $3$. It is pointed out
that this may be the result of a phase transition which has a critical
temperature of roughly a few meV
(in the cosmological units, $~\hbar = c =k = 1$),
which is natural in the context of
massive neutrinos.
In fact,  the neutrino masses required for quasar production
and those required to solve the solar neutrino problem by the MSW mechanism
are consistent with each other. As a bonus,   the cosmological constant
implied by this model may also help resolve the discrepancy between the
recently measured value of the Hubble Constant and the age of the universe.

We illustrate the real time formalism by studying one of the most
important time-dependent and non-equilibrium phenomena associated
with phase transitions.
The non-equilibrium dynamics of the first stage of the reheating process,
 that is dissipation via
particle production is studied in  scalar field theories.
We show that a complete
understanding of the mechanism of dissipation via particle production
requires a non-perturbative resummation.
We then study a Hartree approximation and clearly  exhibit dissipative effects
related to particle production.
The effect of dissipation by Goldstone bosons is
studied non-perturbatively in the
large N limit in an O(N) theory.

We also place our work in perspective and point out some of the
related issues which clearly need further exploration.

\end{quotation}

\newpage

\centerline{\bf Acknowledgements}

I have been very lucky to have wonderful people
close to me. I have benefitted greatly from my interactions with these
people.  I owe a special debt of gratitude to my advisor,
Prof.  Richard Holman.
Rich Holman has the rare quality of combining a great depth of knowledge
with an uncommon ability to communicate.
In many discussions with him I have learnt
a great deal and come to appreciate his great depth of knowledge in many
different fields of Physics. In addition,  I have come to appreciate the
complete seriousness with which he responds to questions which can lead to
a discussion of new ideas.
His encouraging attitude is able to convert questions into ideas and
convert fuzzy ideas into well-directed research.  Thus,  his great depth
of knowledge and his encouraging attitude both help to make him one of
the greatest teachers I have known.
In addition to the regular academic interactions,  through personal
interactions Rich transmits his enthusiasm for physics and for
life.  His cheerful attitude helps those around him keep their spirits up
even when they are temporarily stuck in what looks like a difficult
situation.
I have benefitted immensely in many ways due to my interactions with
Rich.

Dan Boyanovsky has been in many ways a co-advisor to me.
I thank him for his patient explanations,  helpful discussions
and encouraging remarks.  I thank Lincoln Wolfenstein for many
stimulating  and encouraging discussions.  For sharing their
knowledge with me I also thank Ling-Fong Li,  Hector de Vega and Ted Newman.
Posthumous thanks go to Dick Cutkosky;  I benefitted from
interactions with him early in my Ph.D.  and wish I had learned more from him.
Towards the later stages of my Ph.D.  I benefitted from discussions with
David Brahm,  Paul Geiger and Martin Savage.
I thank Bob Kraemer and Helmut Vogel for continuosly pushing me to prove
that cosmology is an experimental science too.
I thank Da-Shin Lee for many fruitful
discussions and frequent collaborations.
Thanks also to Joao Silva,
Dan Cormier,  Prem Sivaramakrishnan and James Walden for the chit-chat
that helped keep the physics in perspective.
In fact,  thanks go to many people at Carnegie-Mellon who helped make the
stay here enjoyable.  I thank Nancy Watson,  Mary Shope and Lori Marcus
for their friendly help and providing me with an unsolved mystery:
where does the bearaucracy hide at Carnegie-Mellon?

I,  of course,  offer special thanks to my family and in particular
to my parents who supported and encouraged me
through my  formative years to the present to do what
I enjoyed doing (but do it well!) !
Thanks most of all to my wife Jaya,  for all the happy times
and also for putting up with my moody and temperamental side
that shows up when the sun stops shining and other such
unforeseen calamities happen.  But mostly it's been happy times!
Thanks for all the happy times!

  \tableofcontents

\def\be{\begin{equation}}
\def\ee{\end{equation}}
\def\ba{\begin{eqnarray}}
\def\ea{\end{eqnarray}}
\def\la{\mathrel{\mathpalette\fun <}}
\def\ga{\mathrel{\mathpalette\fun >}}
\def\fun#1#2{\lower3.6pt\vbox{\baselineskip0pt\lineskip.9pt\ialign{$\mathsurround=0pt#1\hfill##\hfil$\crcr#2\crcr\sim\crcr}}}
\def\re#1{{$^{\ref{#1}}$}}
\def\VEV#1{{{\langle#1\rangle}}}
\def\mpl{{m_{Pl}}}
\def\trec{{{T_{\rm rec}}}}
\def\phit{{{\widetilde{\phi}}}}
\def\ni{\noindent}
\def\non{\nonumber}
\def\eps{\epsilon}
\def\blam{\Lambda}
\def\phif{{\phi/f}}
\def\m{{\cal M}}
\def\mn{m_{\nu}}
\def\htemp{\widehat{T}}
\def\htempc{\widehat{T}_0}
\def\hmr{\widehat{m_r}}
\def\hm{\widehat{\cal M}}
\def\half{{{\frac{1}{2}}}}
\def\dslash#1{#1\!\!\!/\!\,\,}
\def\ApJ{{\it Astrophys.~J.} }
\def\numu{{\nu}_{\mu}}
\def\nuee{{\nu}_e}
\vspace{0.5in}
\chapter{ INTRODUCTION TO FINITE TEMPERATURE FIELD THEORY
AND ITS RELEVANCE TO COSMOLOGY}


\baselineskip=24pt

\vspace{24pt}

\section{Introduction}

Cosmology is one of the most rapidly growing fields of physics today.
This is so because there is a fairly large number of observations which
 we are now in position to make which have been impossible to do so far in
 mankind's history. The combination of both space based observations and
improvements in technology and techniques of ground based observations is
 dramatically improving both the quantity and quality of data now available.
 These data will clearly lead to an improved understanding of the physics
of the universe we live in.

There are many examples of areas of cosmology where  better data continues
to lead to better understanding. In particular, structure fomation continues
to benefit from data from a variety of sources. These data includes improving
determination of the clustering of galaxies and other
larger scale structures\cite{LSS}.
Further light is being shed on structure formation by examining number
densities of quasars\cite{Scmidt},
clustering properties of quasars\cite{Fang} and the study of
Lyman-$\alpha$  clouds\cite{Lya}.
A particularly important piece of information that has
direct implications for structure formation has become available in recent
years - the detection of anistropies in the Cosmic Microwave Backround
Radiation(CMBR)\cite{CMBR}.
This has been measured by COBE and continues to be pursued by
ground based and balloon based experiments. We are starting to get information
 on the CMBR anisotropy on a variety of angular scales and this together with
 available data on the clustering of galaxies and other larger scale
structures is starting to severely constrain models of structure formation.

Another example of fresh observations which have the potential of leading
to an improved understanding of the physics of the universe we live in is
the accurate determination of the Hubble Constant using Cepheid variables
as standard candles\cite{Pierce,Freedman}.
This is among the many achievements of
the new refurbished
Hubble Space Telescope. The ability of the Space Telescope to see at a level
of detail which has not been possible earlier continues to have an impact on
 many fields of astrophysics. Another of its significant contribution has
been to provide compelling evidence for the existence of supermassive
blackholes at the center of the galaxies. An examination of the rotation
velocities of objects close to the center of the galaxy implied the eixstence
of an enormously massive object confined to such a small region of space
that there is no known mechanism of preventing the collapse of such a massive
 object to form a black hole.

The determination of the Hubble Constant is an example where both previously
impossible space based observations\cite{Freedman}
and improvements in ground based
techniques and technology\cite{Pierce}
are happening at the same time to complement and
confirm improved, more accurate results.

The accurate determination of the Hubble constant is also an example of a
recent observation that has important implications for cosmology and
the potential of leading to an improved understanding of the physics of the
universe we live in.
One can use the Hubble
constant to determine the ``expansion age'' of the universe which is the age
of the universe  derived on the assumption that we live in a  flat universe
with no vacuum energy. This expansion age may turn out to be less than the
observed ages of globular star clusters. If all the data stand up to the
 test of time then clearly we have a very  important piece of fresh
information about the universe.

One of the ways of resolving this apparent conflict between the age of the
universe and the Hubble constant is to introduce a cosmologiacal constant
in Einstein's equations which determine the evolution of the universe. The
naturalness or unnaturalness of this solution is discussed in a later
chapter and will be commented on later, but there is a further point about
the role of observations in this context that we would like to make here.

The existence and magnitude of the cosmological constant is an issue which is
also benefitting from continued improvements in astronomical observations.
In particular, since the cosmological constant modifies the geometry of
the universe and so influences the propogation of light it can be
 observationally determined or at least constrained. Thus by observing such
 quantities as quasar lensing frequencies and comparing them to those expected
for various values of the cosmological constant one can determine or at
least constrain the value of the cosmological constant\cite{gravlens}.
There is admittedly a dependence on the modelling of the time evolution
due to ongoing structure formation involved, but for a given detailed
model of structure formation one can use the observations to shed light
on the cosmological constant issue. Indeed, considerable progress has
been made in this field too in the last few years and continued improvements
will only make this a further source of rich information on cosmology.

So far we have given examples of ongoing observations that are yielding
important information about cosmology. There is also a host of observations
that will be coming on-line in the future which will undoubtedly give us
valuable new information about cosmology. In particular,
neutrino astronomy\cite{neuastron}
and gravitational wave astronomy\cite{gravwave}
will allow us to take a look at the universe
in a way we have never been able to explore previously.

It is clear that cosmology is becoming an intensely data-driven subject.
The fresh new data that continues to stream in will undoubtably lead to
an increased understanding of the universe we live in.
However, to use all the data efficiently and effectively, utilize it to
yield insights into the fundamental physics of the processes involved
requires us also to equip ourselves with some theoretical tools and
techniques.

Field theory is clearly one of the essential tools for an understanding
of cosmology. In particular, finite temperature field theory plays a central
role in understanding hot big bang cosmology. This is perhaps underlined
by the recent COBE observation of the perfect thermal Planckian spectrum
for the photon gas that inhabits the universe. The Hubble expansion then
tells us that at earlier times the energy density was higher leading to
higher temperatures for the photon gas at earlier epochs.
Also at the earlier higher densities the higher interaction rates
would imply that there was thermal equilibration of the various particle
distributions with the photon gas at earlier epochs.

In addition to this we believe there were a number of phase transitions
during the evolution of the universe. These phase transitions have
important implications for cosmology. One of these phase transitions is
believed to be responsible for inflation which solves many of the
problems of standard cosmology, such as the flatness problem and the entropy
problem\cite{inflation,reviews}.
The natural tool to study all these phenomena is finite temperature
field theory.
This point has been made in a number of excellent contemporary
books on cosmology\cite{kolbturner,lindebook}.

In this thesis, we  attempt to provide an introduction to finite
temperature field theory geared towards those interested in cosmology.
Towards this goal we have used the techniques of finite temperature field
 theory to address some of the subjects and issues of current interest to
cosmologists.



Phase transitions play an important role in modern theories of particle
physics and cosmology\cite{kolbturner,lindebook}.
Symmetry breaking phase transitions are an
integral part of our current understanding of the unification of
fundamental interactions\cite{li}.




In the Big Bang cosmology
as the universe evolves it expands and it cools. Starting at high
temperatures with a very symmetric universe as the universe
expands and cools it goes through a sequence of phase transitions.
Thus, first the GUT symmetry occurs followed by the electroweak
symmetry breaking followed by other phase transitions.

Among the interesting cosmological issues that are intimately linked
with these phase transitions are the formation and evolution of
topological
defects\cite{kibble,vilenkin,EdCope,turok,branden,boysinlee},
inflation\cite{guth,abstein,lindinf,abpi},
structure formation and baryogenesis.

It is clear that we need a systematic framework in which to address
all of these issues. Finite temperature field
theory\cite{kirlin,doljac,weinberg,coleman,kapusta,rivers}
is definitely the framework one needs to use to get detailed
quantitative information about these interesting and important events.
Within the framework of finite temperature field theory there are two
different formalisms one could use.
These are respectively the imaginary time formalism and the
real time formalism of finite temperature field theory.
Depending on the particular questions or issues one wishes to
address one formalism may be better than the other
and on occasion one of the formalisms may not be relevant to the
issues to be addressed.
In particular, to determine the existence of phase transitions,
the magnitude of the critical temperature and the nature of
topological defects formed at the phase transition one may wish to
use the imaginary time formalism.
However to study the time dependent and non-equilibrium phenomena
associated with phase transitions the natural tool to use is the
real time formalism of finite temperature field theory.
The various uses of the two different formalisms of finite temperature
field theory will be illustrated in detail by studying specific
examples of current interest.

Phase transitions that occur after the decoupling of matter and radiation have
been discussed in the literature as Late Time Phase Transitions (LTPT's). The
original motivation for considering
LTPT's\cite{HillFry} \cite{WASS} \cite{PRS} \cite{FrHW} \cite{GHHK}
\cite{HolSing}was the need to reconcile the
extreme isotropy of the
Cosmic Microwave Background Radiation (CMBR)\cite{CMBRol}
with the existence of large scale structure\cite{LSS} and also the existence of
quasars at high redshifts\cite{quas}.

Discussions of realistic particle physics models capable of generating LTPT's
have been carried out by several authors\cite{GHHK}  \cite{FrHW}. It has been
pointed
out that the most natural class of models in which to realise the idea of
LTPT's are models of neutrino masses with Pseudo Nambu Goldstone Bosons
(PNGB's). The reason for this is that the mass scales associated with such
models can be related to the neutrino masses, while any tuning that needs to
be done is protected from radiative corrections by the symmetry that gave rise
to the Nambu-Goldstone modes\cite{'thooft}.

Holman and Singh\cite{HolSing} studied the finite temperature behaviour of the
see-saw model of neutrino masses and
found phase transitions in this model which
result in the formation of topological defects.  In fact, the critical
temperature in this model is naturally linked to the neutrino masses.

The original
motivation for studying the finite temperature behaviour of the see-saw
model of neutrino masses came from a desire to find realistic particle physics
models for Late Time Phase Transitions. It now appears that this may also
provide a physically appealing and observationally desirable magnitude for
the cosmological constant.

In particle physics one of the standard ways of generating neutrino masses has
been the see-saw mechanism \cite{ramslan}. These models involve leptons
and Higgs fields interacting by a Yukawa type interaction. We computed the
finite temperature effective potential of the Higgs fields in this model. An
examination of the manifold of degenerate vacua at different temperatures
allowed us to describe the phase transition and the nature of the topological
defects formed.

To investigate in detail the finite temperature behaviour of the see-saw
model we selected a very specific and extremely simplified version of the
general see-saw model. However,
we expect some of the qualitative features displayed by our specific
simplified model to be at least as rich as those present in more
complicated models.

In the following chapter we will study in some detail a specific simple
version of the see-saw model of neutrino
masses\cite{ramslan}. By studying the finite
temperature behavior of this model Holman and
Singh\cite{HolSing} demonstrated
the existence of phase transitions, establishing the magnitude of the
critical temperature. This was done by using the imaginary time
formalism of finite temperature field theory. By studying the manifold of
degenerate vacua we also established the nature of topological defects
that would be formed. The phase transition implied by this model may
have interesting cosmological consequences if we use the neutrino
masses implied by the
MSW solution\cite{MSW} to the solar neutrino problem.
In particular, these phase transitions might help explain the peak in
the quasar space densities observed at redshifts between 2
and 3\cite{quasars}.
Also, the magnitude of the cosmological constant implied by this
model might help resolve the apparent conflict between the age of the
universe and the Hubble
constant\cite{cosmoc}. We will study the details of the
finite temperature see-saw model in Chapter 2.

The real time formalism of finite temperature field
theory\cite{smk,hu,rivers}
has been
used by us to study time dependent and non-equilibrium phenomena
during and following a phase transition such as
domain formation and
growth\cite{boysinlee},
thermal activation\cite{bhlss},
dissipation via particle production and reheating\cite{bdhls,dpf}.
In Chapter 3 we will study in detail
the mechanism of dissipation via particle production as an illustration
of the use of the real time formalism to study time dependent
phenomena following a phase transition.

We next
review the formalism we will use in later chapters. The remainder
of this chapter will be devoted to introducing first the imaginary
and then the real time formalisms of finite temperature field theory.

\section{The Imaginary Time Formalism}

In finite temperature field theory the basic quantity of interest
are the finite temperature Green's functions. To study phase transitions
and identify the critical temperature and nature of topological
defects one uses the effective potential which can be expressed
in terms of 1 particle irreducible Greens functions as discussed below.

Let us quickly review the zero temperature field theory and the zero
temperature effective potential first and then discuss their finite
temperature counterparts. We will consider the simplest possible
example of a scalar field theory to introduce the
formalism\cite{li,CW} .

The starting point of our discussion is the generating functional for the
Greens functions,

\be
W[J] = \int D\phi \exp \left\{ i \int d^4 x (L + J \phi ) \right\}
\ee
where $ L $ is the Lagrangian density describing the field theory
of interest, $\phi$ is the scalar field of interest and $J$ is the source
which allows us to express the n-point greens functions in terms
of the generating functional in the following way:

\be
G^{(n)} ( x_1, x_2, ... , x_n) = \frac{1}{i^n} \frac{1}{W} \left[
\frac{\delta^n W[J]}{\delta J(x_1) ... \delta J(x_n)} \right]_{J=0}.
\ee

Similarly we can  introduce the quantity, $ Z[J] \equiv - i~ \ln W[J] $,
which can be shown to be the generating functional of connected
greens functions\cite{li},

\be
G_c^{(n)} ( x_1, x_2, ... , x_n) = \left[ \frac{\delta^n \ln W[J]}{\delta
J(x_1) ... \delta J(x_n)} \right]_{J=0}.
\ee

Then, the classical field which is the vacuum expectation value of the field
is given by,

\be
\phi_c(x) \equiv \frac{\langle 0 | \phi(x) | 0 \rangle}{\langle 0  | 0 \rangle}
= \frac{\delta Z[J]}{\delta J(x)}.
\ee

The effective action, $\Gamma$ is the functional Legendre transform of the
generating functional $Z[J]$,

\be
\Gamma[\phi_c] = Z[J] - \int d^4 x J(x) \phi_c(x) .
\ee

One can also expand $\Gamma[\phi_c]$ in powers of $\phi_c$ as,

\be
\Gamma[\phi_c] = \sum_n \frac{1}{n!} \int d^4 x_1 ... d^4 x_n
\Gamma^{(n)}(x_1,x_2, ... x_n) \phi_c(x_1) ...  \phi_c(x_n).
\ee

It can be shown that,
$  \Gamma^{(n)}(x_1,x_2, ... x_n) \equiv \frac{\delta^n \Gamma}{\delta
\phi_c(x_1) ... \delta \phi_c(x_n)} $ are the 1 particle irreducible (1PI)
n-point
greens functions. Recall that the 1PI Greens functions receive
contributions only from Feynman diagrams which cannot be disconnected
by cutting any one internal line.

Making a derivative expansion of the effective action allows us to isolate
the effective potential as the lowest order term in the expansion,

\be
\Gamma[\phi_c] =  \int d^4 x \left[ - V_{eff}(\phi_c) + \frac{1}{2}
(\partial_{\mu} \phi_c)^2 F(\phi_c) + ... \right].
\ee

In particular, for $ \phi_c(x) =  \phi_c = constant $, we have
$ \partial_{\mu} \phi_c = 0  $. Thus, by taking Fourier transforms one obtains
from the previous equations the following relation,

\be
V_{eff}(\phi_c) = -  \sum_n \frac{1}{n!}  \Gamma^{(n)}(p_1 = 0, ... p_n = 0)~
\phi_c^n .
\ee


\subsection{Calculation of the effective potential using the tadpole method}

The effective potential formalism is the most well-known and widely
used formalism to study the phenomenon of spontaneous symmetry breaking
in field theory. The formalism is by now well-developed and there
exist a number of excellent discussions on the effective potential
and methods to calculate it\cite{CW,li}. The most straightforward and
frequently discussed method of calculating the effective potential
is by evaluating all the 1PI n-point Greens function and then
explicitly performing
the sum over n as required by the relation given earlier to
obtain an expression for the effective potential.
There exists another method which is less well-known, a little devious,
usually easier to implement in practice and is our method of choice to
calculate the effective potential. By being a little devious we can express
the effective potential completly in terms of a 1-point 1PI Greens functions
and this makes the computaion of effective potential much simpler in practice.
We now turn to a description of this method and its illustration using
a simple example.

By introducing a shift in the fields,
\be
\phi  \rightarrow \phi - \omega
\ee
one can obtain an expression for the effective potential in terms of the
1PI n-point functions in the shifted theory,
$ \Gamma_{\omega}^{(n)} $ which is more useful for
actually calculating the effective potential in practice. Thus,

\be
V_{eff}(\phi_c) = -  \sum_n \frac{1}{n!}  \Gamma_{\omega}^{(n)}(p_1 = 0, ...
p_n = 0)~ (\phi_c - \omega)^n .
\ee

This implies,

\be
\left[ \frac{d V_{eff}}{d \phi} \right]_{\phi = \omega} = -
\Gamma_{\omega}^{(1)}(p_1 = 0).
\ee

Therefore, we have,

\be
V_{eff} = - \int  \Gamma_{\omega}^{(1)} d \omega.
\ee

Note that by making the above transformations we can calculate the effective
potential by knowing only the 1-point 1 PI function in the shifted theory
as opposed to calculating all the n-point functions and then summing over
n in the original expression for the effective potential. This method is
called the tadpole method because the calculation of the 1PI 1-point
function,
$  \Gamma_{\omega}^{(1)} $ involves the computation of a Feynman diagram
which bears a closer resemblance to a tadpole than the penguin diagram's
resemblance to a penguin.

Let us illustrate the method with a simple example. Consider a scalar
field theory with the following tree level potential,

\be
V_0(\phi) = \frac{1}{2} m^2 \phi^2 + \frac{\lambda}{4!} \phi^4.
\ee

Performing the shift, $ \phi  \rightarrow \phi - \omega $, we get,

\be
V_{\omega} = \frac{1}{2} m^2 ( \phi - \omega)^2 + \frac{\lambda}{4!} ( \phi -
\omega)^4.
\ee

In the shifted theory the mass of the $\phi$ field, determined from the
coefficient of the  $\phi^2$ term is given by,
$ M_{\phi}^2(\omega) \equiv m^2 +  \frac{\lambda  \omega^2}{2}  $.
The computation of $ \Gamma_{\omega}^{(1)} $  involves the cubic
self-coupling i.e. the coefficient of the $\phi^3$ term which is
$ \frac{\lambda}{3!} \omega $. Thus, the 1 loop contribution to
 $ \Gamma_{\omega}^{(1)} $ is,

\ba
\Gamma_{\omega}^{(1)}(p_1 = 0) &  = &  - \frac{i}{2} \int \frac{d^4 k}{(2
\pi)^4} \frac{\lambda \omega }{k^2 -  M_{\phi}^2(\omega) + i \epsilon} \\
& = & - \frac{d V_{eff}}{d \omega}. \nonumber
\ea

Thus,

\ba
V_{eff} & = & - \int  \Gamma_{\omega}^{(1)} d \omega \nonumber \\
& = & \frac{i}{2} \int  \frac{d^4 k}{(2 \pi)^4} \ln \left[ \frac{k^2 -
M_{\phi}^2(\omega) + i \epsilon}{\Lambda^2} \right].
\ea

By performing the rotation to Euclidean space the above integral
can be evaluated to yield the 1 loop contribution to the effective
potential for this theory  to be,

\be
V_{eff} = \frac{1}{64 \pi^2} \left\{ - \frac{\Lambda^4}{2} + 2 \Lambda^2
M_{\phi}^2(\omega = \phi_c) +  M_{\phi}^4(\omega = \phi_c) \left[ \ln \left(
\frac{ M_{\phi}^2(\omega = \phi_c)}{\Lambda^2} \right) - \frac{1}{2}
\right]\right\}
\ee
where $\Lambda$ is the momentum cutoff introduced to regularise the integral.
This theory can now be renormalized by adding counterterms and
specifying a renormalization prescription as is done below.
The counterterms we add are
\be
\delta V_{ct} =  - \frac{1}{2} \delta m^2 \phi_c^2 - \frac{\delta \lambda}{4!}
\phi_c^4 + \delta V.
\ee

The total potential,$V$ then becomes the sum of the bare potential and
$\delta V_{ct}$.We specify the following renormalization prescription:

\be
\left[ \frac{d^2 V}{d \phi^2} \right]_{\phi=0} = m_r^2 \; , \; \left[ \frac{d^4
V}{d \phi^4} \right]_{\phi=0} = \lambda_r \; , \;V(\phi = 0) = V_o.
\ee

This yields the following expressions for the renormalized parameters:

\ba
V_o & = &\delta V + \frac{m^4}{32 \pi^2} \left[ \ln \left( \frac{m^2
}{\Lambda^2} \right) - \frac{1}{2} \right] \\
m_r^2 & = & \delta m^2 + \frac{\lambda \Lambda}{32 \pi^2} + m^2 + \frac{\lambda
 m^2}{32 \pi^2} \ln \frac{m^2}{\Lambda^2} \\
\lambda_r & = & \lambda + \delta \lambda  \frac{3 \lambda^2}{32 \pi^2} \left[
\ln \left( \frac{m^2 }{\Lambda^2} \right) + 1 \right]. \\
\ea

Thus in terms of these renormalized parameters the complete effective
potential becomes,
\be
V_{renorm} =  \frac{1}{2} m_r^2 \phi^2 + \frac{\lambda_r}{4!} \phi^4 +
\frac{m_{\phi}^4}{64 \pi^2} \left[ \ln \left( \frac{m_{\phi}^2 }{\Lambda^2}
\right) - \frac{1}{2} \right] + V_o
\ee
where $m_{\phi}^2 = m_r^2 + \frac{\lambda_r \phi^2}{2}$.

In equilibrium
finite temperature field theory we are interested in evaluating the
finite temperature Green's functions,

\be
G_T^{(n)}(x_1,x_2, ... x_n) = \frac{Tr[ \phi(x_1) \phi(x_2) ...  \phi(x_n) e^{-
\beta H} ] }{Tr[ e^{- \beta H} ]}.
\ee

Recall that $\beta = \frac{1}{k T}$, where $T$ is the temperature.
The temperature is a well defined quantity only for systems in equilibrium.
In cosmology, we are frequently interested in systems which are in
local thermal equilibrium. This happens when the interaction rates are
high compared to the expansion rate of the universe.
In such situations the above quantity is meaningful and of interest.
There are of course, other situations such as when a phase transition
occurs that the system is far from equilibrium and we will later
discuss a formalism which allows us to keep track of the time
evolution under these circumstance. For now, let us concentrate
on equilibrium finite temperature field theory.
The fundamental quantity of interest for equilibrium
finite temperature field theory
is the partition function which is
the generating function of Green's functions,

\ba
Z_{\beta} & = & Tr[ e^{- \beta H} ] \\
& = & \int D \phi \langle \phi |  e^{- \beta H} | \phi \rangle .
\ea

Thus we need to evaluate
$  \langle \phi |  e^{- \beta H} | \phi \rangle $. Recall that similar
quantities,

\be
\langle q'; t' | q ; t \rangle =   \langle q' |  e^{- i H (t-t')} | q \rangle
\ee
are frequently evaluated in quantum mechanics. This points the way to
using well developed techniques in zero temperature field theory for
calculations in finite temperature field theory by making the identification
$ i \tau = - \beta $. Thus the inverse temperature is formally identified
with the imaginary time direction and it is for this reason that the
formalism described in this section is called the imaginary time
formalism of finite temperature field theory.

An important point to notice is that since we are interested in evaluating
the trace, the ` initial ' and ` final ' states have to be identified
leading to periodicity in imaginary time $\tau$ for bosons. Fermions
are anti-symmetric under the exchange of states and this leads to
an anti-periodicity in imaginary time for the fermions.

Knowing the Hamiltonian allows one to extract the finite temperature
Feynman rules which are then used to calculate the finite temperature
effective potential using the tadpole method. Let us do this in some detail
for the $\lambda \phi^4$ theory.
The most important modification to the Feynman rules at finite
temperature is in the propogator.
In fact, the factors at the verices and the combinatorics
are the same at finite temperature as they are at zero temperature.
The important modification
of the propogator comes essentially from the fact that we have
definite periodicity in imaginary time for the bosons and anti-periodicity
in imaginary time for the fermions.
To determine the propogator it is sufficient to consider only the
quadratic part of the  action,

\be
S_0[ \phi ] = \int_0^{\beta} d \tau d^3 x ~~ \phi \left[ -
\frac{\partial^2}{\partial \tau^2} - \nabla^2 + M^2 \right] \phi.
\ee

Introducing the Fourier expansion for the field,

\be
\phi(x,\tau) = \frac{1}{\beta} \sum_{m = - \infty}^{ \infty} \int
\phi_m(\underline{k} ) e^{i(\underline{k} \cdot \underline{x} + \omega_{2m}
\tau )} d^3 k
\ee
one gets the propogator in momentum space to be,

\be
D(\underline{k}, \omega_{2m}) = \frac{1}{ \omega_{2m}^2 + k^2 + M^2}
\ee
where the quantity $ \omega_{2m} = \frac{2 \pi m}{\beta}$ results from
the periodicity in imaginary time. Also notice that the periodicity
in imaginary time leads to a dicrete sum in the fourier transform.
Thus,

\be
\int \frac{d^4 k }{(2 \pi)^4} \rightarrow \frac{1}{\beta} \sum_m \int \frac{d^3
k }{(2 \pi)^3}.
\ee

For fermions we have anti-periodicity in imaginary time  and thus instead
of $ \omega_{2m}$, the quantity entering expressions is,
$ \omega_{2m+1} =  \frac{ \pi (2m+1)}{\beta} $.

Once again we will use the tadpole method, this time to compute the
finite temperature correcions to the effective potential.
The relation which gives the finite temperature correction, $\Delta V_T$ is,

\ba
\frac{d \Delta V_T}{d \omega} & = & \frac{1}{\beta} \sum_m \int \frac{d^3 k
}{(2 \pi)^3} \frac{\lambda \omega }{ \omega_{2m}^2 + k^2 + M_{\phi}^2(\omega) }
\\
& = &  \frac{1}{\beta} \sum_m \int \frac{d^3 k }{(2 \pi)^3} \ln \left[ \frac{
\omega_{2m}^2 + k^2 + M_{\phi}^2(\omega) }{\Lambda^2 } \right].
\ea

So far, the only thing that is really different from the zero temperature
discussion has to do with the fact that there is a periodicity in imaginary
time leading to a discrete fourier sum in the timelike direction.
To proceed further, we choose to evaluate the sum first and then evaluate
the integral.
The sum can be evaluated by using the identity,
\be
\sum_{m=-\infty}^{\infty} \ln ( \omega_{2m}^2 + E_k^2 ) = \beta E_k + 2 \ln (1
- e^{-\beta E_k} ) + (E_k~ independent~ terms)
\ee
where
\be
E_k^2 =  k^2 + M_{\phi}^2 \; , \;   \omega_{2m} = \frac{2 \pi m}{\beta}.
\ee

Using the above identity we can evaluate the finite temperature correction
to the effective potential,
\be
\Delta V_T =  \int \frac{d^3 k }{(2 \pi)^3} \frac{1}{2} E_k + \int \frac{d^3 k
}{(2 \pi)^3} \ln \left[ 1 - e^{- \beta E_k} \right].
\ee
At this point we note that the second term on the right is finite and the
first term on the right hand side contians all the divergent part.
Further, we note that in fact the divergence structure of this first
term is identical to the divergence structure of the zero-temperature
part of the effective potential. It thus follows
that the  counterterms that we introduced to absorb the divergences of the
zero-temperature effective potential will be sufficient to absorb all the
divergences of the finite temperature corrections to the effective
potential. Hence, the net effect of the finite temperature corrections
is to add only the second term on the right hand side
(which we will call $\Delta V_T^f$ to the expression
for the renormalised effective potential. Let us evaluate the
consequences of this term in some further detail.
First, let us set $ x = \beta k$ and do the solid angle integral.
This yields,
\be
\Delta V_T^f = \frac{1}{2 \pi^2 \beta^4} \int dx~x^2 \ln \left[ 1 - e^{-(x^2+
\beta^2 M^2(\phi_c))^\frac{1}{2}} \right].
\ee

For the region where $\frac{M}{T} << 1$ we can do a high temperature
expansion to the above integral which yields,
\ba
\Delta V_T^f & = & \frac{-\pi^2}{90} T^4 + \frac{ M^2(\phi_c) T^2}{24} -
\frac{T}{12 \pi} \left( M^2(\phi_c) \right)^{3/2}  \nonumber \\
&   &  - \frac{1}{64 \pi} M^4(\phi_c) \ln \left[ \frac{ M^2(\phi_c)}{T^2}
\right] + \frac{(3/2 - 2 \ln 4 \pi - 2 \gamma )}{64 \pi^2}  M^4(\phi_c)
\nonumber \\
+ O(M^6 \beta^2).
\ea

To lowest order,
\be
\Delta V_T^f = \frac{-\pi^2}{90} T^4 + \frac{ M^2(\phi_c) T^2}{24}.
\ee
Thus, if we collect the coefficient of the $\phi_c^2$ from the zero temperature
potential and the finite temperature correction and define this coefficient
as $m^2(T)$, then we get,
\be
m^2(T) = m_{T=0}^2 + \frac{\lambda T^2}{24}
\ee
which indicates how a theory which has a broken symmetry at zero temperature
($m_{T=0}^2 < 0 $) can have the symmetry restored at some higher temperature
such that $m^2(T) = m_{T=0}^2 + \frac{\lambda T^2}{24} > 0 $.
Thus, the critical temperature will be given by
$m^2(T_c) = 0 = m_{T=0}^2 + \frac{\lambda T_c^2}{24} $.

Thus we see how the imaginary time formalism of finite temperature
field theory can be used to determine the critical temperature given
a model of the interactions of fields we are interested in, in terms
of the parameters of the model. In this section we have used the simplest
possible model of interactions to introduce the concepts, formalism and
techniques of the imaginary time formalism of finite temperature field
theory. In chapter 2 we will use this same formalism and these same
techniques to study the finite temperature behaviour of a specific
see-saw model of neutrino masses. As we will see the finite temperature
behaviour of this model may have many interesting cosmological
consequences including a  possible explanation of the peak in
comoving quasar space densities as well as providing a cosmological
constant of the right magnitude to remove the discrepancy between
the Hubble constant and the age of the universe. Of course, in addition
to being of current interest this model will also allow us to
illustrate the techniques and utility of the imaginary time formalism of
finite temperature field theory. In particular, we will be able to analyze
the manifold of degenerate vacua of the effective potential in that model
to extract information about the nature of topological defects that
will be formed at various temperatures.

\section{ The Real Time Formalism }

So far we have studied the imaginary time formalism of finite temperature
field theory. As already pointed out however, the imaginary time formalism
cannot be used to  study time dependent phenomena
since the ` imaginary time ' is identified with temperature. The best
one can hope to do in this formalism is to stipulate that the temperature
is some specified function of time $T(t)$. However, this implies that the
system is always very close to some equilibrium configuration with a well
defined temperature $T$. In particular, this implies that the density
matrix is always expressible in the form $ \rho = e^{-\beta H} $. As we shall
see during the transition period immediately following a phase transition
the distribution function does not look like it can be fit by a purely thermal
distribution function with a single well defined temperature at each
instant of time.

The physical quantity best suited to describe the time evolution of a
quantum statistical system out of equilibrium is the density matrix,
$\rho(t)$.
The density matrix can be expressed in terms of a complete set of eigenstates
$|\psi_i \rangle$ in the following way,
$\rho = \sum_i |\psi_i \rangle p_i \langle\psi_i | $
where $p_i$ represents the probability of finding the state of the system
in the state $|\psi_i \rangle$. Note that the density matrix need not
neccesarily be diagonal, however it must be hermitian and hence can be
diagonalised by a unitary transformation and expressed in the form given above.
The density matrix is particularly useful for describing the state of a system
which is not a pure quantum mechanical state but is instead in a mixed
state. In such cases the only information we may possess is the statistical
information contained in the probabilities of finding the various different
eigenstates of the system. An example of such a mixed statistical system
is completely unpolarised light for which the density matrix would be
an equally weighted sum over the different polarization states of light.
Another example of a mixed statistical system which occurs frequently
in nature and will be of concern to us is a system in thermal equilibrium
at a given temperature $T$. For such a system all that we may know is that
the probability of finding an eigenstate with energy $E_j$ is proportional to
$e^{-\frac{E_j}{k T}}$. For statistical systems we are frequently interested
in the time evolution of a specific initial mixed state under the influence
of a time dependent Hamiltonian.

The time evolution of the system is then determined by the quantum
Liouville equation,

\be
i \hbar \frac{\partial \rho}{\partial t} = [ H(t), \rho (t) ]
\ee
where $ H(t) $ is the Hamiltonian of the system.

The formal solution of the quantum Lioville equation is given by,

\be
\rho (t) = U(t,t_0) \rho (t_0) U^{-1} (t,t_0)
\ee
where $ \rho (t_0) $ is the intial density matrix at time $ t_0$ and
$ U(t,t_0) $ is the time evolution operator from $t_0$ to $t$.

Typically we are interested in systems that are initially in thermodynamic
equilibrium with a well defined initial inverse temperature
$\beta_i = \frac{1}{k_B T_i}$. The initial density matrix is then given by,

\be
\rho(t_0) = e^{- \beta_i H_i}.
\ee

Notice that if we are interested in the ground state of the system which
is the equilibrium state at zero temperature, we can isolate it by taking
the limit $\beta_i \rightarrow \infty$.

The initial Hamiltonian $H_i$
which governs the behavior of the system for times $t \leq t_0$ prepares the
system to be in equilibrium with a well defined temperature and $\beta_i$ uptil
$t_0$. Interesting non-equilibrium phenomena will occur if for times
$t > t_0$, the Hamiltonian becomes
$H_{evol} (t)$ which is different from the initial
Hamiltonian $H_i$. The time evolution of the expectaion value of any
operator $\cal O$ is given by,

\be
\langle {\cal O} \rangle (t) = \frac{Tr [  e^{- \beta_i H_i}  U^{-1} (t) {\cal
O} U(t) ]}{Tr [  e^{- \beta_i H_i}]}.
\ee

The above expression can be put into a more compact form by noting once
again that the time evolution operator, $U$ also involves the exponential
of the Hamiltonian. In particular, for any time $T < t_0$,

\be
U(T) = \exp [-i T H_i] .
\ee

We can use this fact to absorb the $ \exp [- \beta_i H_i]$ factor into
a modified time evolution operator. Let us re-express,
$\exp [ - \beta_i H_i] = \exp [ -i H_i (T-i\beta_i-T)] = U(T-i\beta_i,T) $.
Further, we insert the factor $U^{-1}(T)U(T) = 1 $ in the numerator and
using the fact that $U^{-1}(T)$ commutes with $\exp [ - \beta_i H_i]$
we can re-express,

\be
\langle {\cal O} \rangle (t) = \frac{Tr [ U(T - i\beta_i,t) {\cal O} U(t,T)
]}{Tr [ U(T - i\beta_i,t)]}.
\ee

We have used slightly convoluted means to arrive at a compact result. In
particular, we have hidden the initial density matrix but we have entered
the complex time plane. Notice that in the imaginary time formalism we
had only the imaginary time direction playing any role. In the real time
formalism both the imaginary direction as well as the real direction
enter the discussion. Thus the real time formalism should really be
called the complex time formalism. However, that might have made the
real time formalism sound even more dauntingly complex than it
already appears. This may have impeded  progress in this field for
a few more years in real time.


We will use the real time formalism to study in detail the mechanism
of dissipation via particle production\cite{bdhls,dpf} in a
later chapter.
This is a process that is very important for understanding reheating
following inflation.
In chapter  3 we use the real time formalism to study in detail one of the
most important time dependent and non-equilibrium phenomena associated
with phase transitions. This is the phenomenon of dissipation of the
initial vacuum energy of the field. The physical mechanism for this
is the particle production which takes place as the field oscillates
about the minimum of its potential. We used the quantum Lioville equation
to study the time evolution of the system starting with an initial
configuration described by a density matrix, $\rho_{initial}$.

Once again the motivation for presenting this discussion in detail is twofold.
First, it is a subject of current cosmological interest in its own right.
Secondly it provides a detailed illustration of the techniques that will
undoubtably be useful in the study of a number of other time-dependent
and non-equilibrium phenomena.




\def\be{\begin{equation}}
\def\ee{\end{equation}}
\def\ba{\begin{eqnarray}}
\def\ea{\end{eqnarray}}
\def\bma{\begin{array}{c}}
\def\ema{\end{array}}
\def\la{\mathrel{\mathpalette\fun <}}
\def\ga{\mathrel{\mathpalette\fun >}}
\def\fun#1#2{\lower3.6pt\vbox{\baselineskip0pt\lineskip.9pt
        \ialign{$\mathsurround=0pt#1\hfill##\hfil$\crcr#2\crcr\sim\crcr}}}
\def\re#1{{$^{\ref{#1}}$}}
\def\VEV#1{{{\langle#1\rangle}}}
\def\mpl{{m_{Pl}}}
\def\trec{{{T_{\rm rec}}}}
\def\phit{{{\widetilde{\phi}}}}
\def\ni{\noindent}
\def\non{\nonumber}
\def\eps{\epsilon}
\def\blam{\Lambda}
\def\phif{{\phi/f}}
\def\m{{\cal M}}
\def\Mu{{\cal M}_{\nu}}
\def\mn{m_{\nu}}
\def\htemp{\widehat{T}}
\def\htempc{\widehat{T}_0}
\def\hmr{\widehat{m_r}}
\def\hm{\widehat{\cal M}}
\def\half{{{\frac{1}{2}}}}
\def\dslash#1{#1\!\!\!/\!\, \, }
\def\ApJ{{\it Astrophys.~J.} }
\def\numu{{\nu}_{\mu}}
\def\nuee{{\nu}_e}
\null\vspace{-62pt}
\vspace{0.5in}
\chapter{IMAGINARY TIME APPLICATION: A LATE TIME PHASE TRANSITION
LINKED WITH MASSIVE NEUTRINOS}
\vspace{.5in}


\vspace{24pt}

\section{Introduction}

One application of the imaginary time formalism that is of current
cosmological interest is to the treatment of Late Time Phase
Transitions(LTPT's).
We will first describe what LTPT's are and then point out why we think
they have a number of extremely interesting and potentially
important cosmological consequences.   Once this is done we will
turn to a detailed discussion of a compelling particle physics
model for such phase transitions.   The detailed investigation of this
model will also illustrate some of  the strenghts of the imaginary time
formalism for determining the existence of phase transitions,   the
critical temperature of the phase transition,  the existence and nature
of topological defects formed during the phase transition in a given
model for interacting fields.

Phase transitions that occur after the decoupling of matter and radiation have
been discussed in the literature as Late Time Phase Transitions.   The
original motivation for considering
LTPT's\cite{HillFry,WASS,PRS,frhw,GHHK,HolSing}was the need to reconcile the
extreme isotropy of the Cosmic Microwave Background Radiation (CMBR)\cite{CMBR}
with the existence of large scale structure\cite{LSS} and also the existence of
quasars at high redshifts\cite{quas}.  The hope that LTPT's would not
disturb the CMBR was not realised\cite{TurnWW}.
In light of the fact that the
anisotropy measurements of the CMBR are getting better\cite{SSM},
the distortions of the
CMBR predicted by  LTPT's may become important observational tests of these
models.

In our mind the current motivation for a continued interest in LTPT's
is twofold.   First,  LTPT's may provide an explanation for the peak observed
in comoving quasar space densities plotted as a function of
redshift\cite{Scmidt}.   As we will see in detail in the discussion later,
the most compelling particle physics model for LTPT's is the
finite temperature behaviour of a see-saw model of neutrino masses.
In such a model one finds a critical temperature which is a few
times the relevant light neutrino mass.   If we input into this
model the neutrino masses required to solve the solar neutrino model
by the MSW mechanism\cite{mswtheory,mswexpt}
 we obtain that the phase transition happens
at the right epoch to explain the peak in quasar space densities.
A second important motivation for a continued interest in the LTPT
model being presented in this chapter is that it may help
remove the discrepancy between the recently determination of the
Hubble constant and the age of the universe.

LTPT's are used to generate density fluctuations either by bubble nucleation
\cite{WASS},  dynamics of a slow rolling field\cite{PRS}
or by the formation of
topological defects such as domain walls(HSF).
The appeal of LTPT's for structure formation lies in the fact that the
density perturbations created are immediately non-linear and so will lead to
much faster formation of structure.
The models we will be
considering in detail in this chapter involve the formation of topological
defects as a result of LTPT's.  The original idea of the production of soft
domain walls as a result of an LTPT (HSF) has received several criticisms.  A
problem with the use of domain walls to form Large Scale Structure has been
pointed out by Kolb and Wang\cite{Rockyun}.  They argue that domain walls may
never form in a LTPT since thermal effects may not be able to drive
different regions of the Universe to different parts of the disconnected vacuum
manifold.
However they also point out that domain walls would
form if large fluctuations in the scalar field exist before the phase
transition.
One of the main critcisms of soft domain walls is that they may produce
significant distortions of the CMBR.  The most significant microwave distortion
comes from collapsing domain wall bubbles\cite{TurnWW}.
These distortions produce
hot and cold spots on $ \sim 1^o$  angular scales and provide a signature
for the formation of domain walls after recombination.

Since the collapse of closed domain wall bubbles to form black holes which
then
act as the central engine of quasars is one of the main themes of this
part of this thesis,
we now turn to a quick discussion of the distortions of the CMBR
this produces.  The temperature shift due to a photon
traversing a collapsing domain wall bubble is
\be
\frac{\Delta T}{T} = 2.  64 \times 10^{-4} h^{-1} \beta A \sigma / (10 MeV^3)
\ee
where $h,~ A,~ \beta$ are dimensionless numerical constants of order unity
and $\sigma$ is the surface tension of the domain wall.   The present
measurements of the CMBR anisotropy then imply\cite{sclu} that
$\sigma < 0.5 ~MeV^3 $.   This constraint,  though important to keep in
mind is not a
problem for the viability of the model being presented in this chapter.
We will return to this point after discussing our particle physics model
for the LTPT.

\section{Particle physics models for LTPT's}

Discussions of realistic particle physics models capable of generating LTPT's
have been carried out by several authors\cite{GHHK,frhw}.  It has been
pointed
out that the most natural class of models in which to realise the idea of
LTPT's are models of neutrino masses with Pseudo Nambu Goldstone Bosons
(PNGB's).  The reason for this is that the mass scales associated with such
models can be related to the neutrino masses, while any tuning that needs
to be
done can be protected from radiative corrections by the symmetry that gave rise
to the Nambu-Goldstone modes\cite{'thooft}.

Holman and Singh\cite{HolSing} studied the finite temperature behaviour of the
see-saw model\cite{ramslan} of neutrino masses
and found phase transitions in this model which
result in the formation of topological defects.   In fact, the critical
temperature in this model is naturally linked to the neutrino masses.

To investigate in detail the finite temperature behaviour of
models allowing for natural LTPT's
we selected a very specific and extremely simplified version of the
general see-saw model.  However,
we expect some of the qualitative features displayed by our specific
simplified model to be at least as rich as those present in more
complicated models.

We chose to study the $2$-family neutrino model.  Because of the mass hierarchy
and small neutrino mixings\cite{mswtheory}  the expectation was
to capture some of the
essential physics of the $\nuee$-$\numu$ system in this way.  The $2$-family
see-saw model we consider requires 2 right handed neutrinos $N_R^i$ which
transform as the fundamental of a global $SU_R(2) $ symmetry.  This symmetry is
implemented in the right handed Majorana mass term by the introduction of a
Higgs field $\sigma_{ij}$, transforming as a symmetric rank $2$ tensor under
$SU_R(2)$ (both $N_R^i$ and $\sigma_{ij}$ are singlets under the standard
model gauge group).  The spontaneous breaking of $SU_R(2)$ via the vacuum
expectation value (VEV) of $\sigma$ gives rise to the large right handed
Majorana masses required for the see-saw mechanism to work.  Also, the
spontaneous
breaking of $SU_R(2)$ to $U(1)$ gives rise to 2  Nambu Goldstone Bosons.
The $SU_R(2)$ symmetry is explicitly broken in the Dirac sector of
the neutrino mass matrix, since the standard lepton doublets $l_L$ and the
Higgs doublet $\Phi$ are singlets under $SU_R(2)$.  It is this explicit
breaking that gives rise to the potential for the Nambu Goldstone modes
via radiative corrections due to fermion loops.  Thus, these modes become
Pseudo
Nambu Goldstone Bosons (PNGB's).

The relevant Yukawa couplings in the leptonic sector are:
\be
-{\cal L}_{\rm{yuk}} = y_{ai} \overline{l_L^a} N_R^i \Phi + y \overline{N_R^i}
N_R^{j\ c}
\sigma_{ij} + \rm{h.  c.  }
\ee
where $a,  i,  j =1,  2$.   The $SU_R(2)$ symmetry is implemented as follows:
\ba
N_R^i & \rightarrow & U^i_j\ N_R^j \nonumber \\
\sigma_{ij} & \rightarrow & U^k_i \sigma_{kl} (U^T)^l_j
\ea
where $U^i_j$ is an $SU_R(2)$ matrix.
The first (Dirac) term breaks the symmetry explicitly.

We can choose the VEV of $\sigma$ to take the form\cite{LI}:
$\langle \sigma_{ij} \rangle = f \delta_{ij}$,  thus breaking $SU_R(2)$
spontaneously down to the $U(1)$ generated by $\tau_2$ (where $\tau_i$ are the
Pauli matrices).  This symmetry breaking gives rise to the $2$ PNGB's, whose
finite temperature effective potential is of interest.

After the Higgs doublet acquires its VEV,  we have the following mass terms
for the neutrino fields:
\be
-{\cal L}_{\rm{mass}} = m_{ai} \bar{\nu}_L^a N_R^i + M \overline{N}_R UU^T
N_R^c + \rm{h.  c.  }
\ee
where $\nu_L^a$ are the standard neutrinos,  $m_{ai} = y_{ai}\ v/{\sqrt{2}}$,
$M = y f/\sqrt{2}$.


 Examining the expression for $-{\cal L}_{\rm{mass}}$ above we notice that
 it may be expressed more compactly by introducing the fields,

 \be
 N_R ' = U^{\dagger} N_R
 \ee

 Then the expression for  $-{\cal L}_{\rm{mass}}$ becomes,

 \be
 -{\cal L}_{\rm{mass}} =  m_{ai} \bar{\nu}_L^a U^i_{j} N_R'^j + M
\overline{N}_R'N_R'^c + \rm{h.  c.  }
 \ee

 At this point we can introduce the $\nu$ - mass matrix to rewrite the above
 terms in a block matrix form :

\ba
-{\cal L}_{\rm{mass}} = \pmatrix{   \bar{\nu}_L &  \overline{N}_R \cr}
\pmatrix{0        & m U \cr (m U)^T  & M \cr} \pmatrix{\nu_L^c \cr N_R^c \cr}
+ \rm{h.  c.  }
\ea

 where the mass matrix,  $ \Mu $,

 \be
 \Mu = \pmatrix{0        & m U \cr (m U)^T  & M \cr}
 \ee

 can be re-expressed in the suggestive form,

 \be
 \Mu = \pmatrix{0 & 0 \cr 0  & M \cr} + \pmatrix{0  & m U \cr (m U)^T  & 0 \cr}
 \ee

 In particular,  in the physical case where $|m_{ai}|<<M$ we can use
 perturbation theory techniques to diagonalise the mass matrix.
 Specifically,  we need to find the eigenvalues of the mass matrix by
 using the second matrix as a perturbation on the first matrix in the
 expression above.   Thus,  the zeroth order eigenvalues are
 $ 0,  0,  M,  M$ and the corresponding eigenvectors can be taken to be
 of the block form
 $\pmatrix{e_1 \cr 0},  ~ \pmatrix{e_2 \cr 0}, ~ \pmatrix{0 \cr e_1},
  ~ \pmatrix{0 \cr e_2}$ where
 \be
 e_1 = \pmatrix{1 \cr 0} \; , \; e_2 = \pmatrix{0 \cr 1}
 \ee

 The most significant effects due to higher order corrections will
 clearly be to the first two eigenvalues which are both 0 at the
 zeroth order in perturbation theory.   Let us calculate the corrections
 to these two eigenvalues explicitly.
 Call the perturbation matrix $M_1$ i.  e.

 \be
 M_1 =  \pmatrix{0  & m U \cr (m U)^T  & 0 \cr}
 \ee

 Then,  the first order corrections to the eigenvalues $\lambda_i$
 for $i = 1, 2$ is given by,

 \ba
 \lambda_i^{(1)} & = & \langle i | M_1 | i \rangle \\
	   & = & \pmatrix{e_i & 0 \cr} \pmatrix{0  & m U \cr (m U)^T  & 0 \cr}
\pmatrix{e_i \cr 0} \\
	   & = & 0
 \ea

 Thus the first non-zero correction to these eigenvalues occur at second
 order in perturbation theory which is given by,

 \be
 \lambda_i^{(2)}  = \sum_{j \neq i} \frac{ \langle i | M_1 | j \rangle  \langle
j | M_1 | i \rangle}{\lambda_i^{(0)}-\lambda_j^{(0)} }
 \ee

 Substituting the explicit form for $M_1$ and the eigenvectors in the
 above expression we get,

 \be
 \lambda_1^{(2)} = - \frac{1}{M} \left[ (m U)_{11}^2 +  (m U)_{12}^2 \right]
 \ee

 and

 \be
 \lambda_2^{(2)} = - \frac{1}{M} \left[ (m U)_{21}^2 +  (m U)_{22}^2 \right]
 \ee

 where

 \ba
 (m U)_{aj} & = & m_{ai} U^i_j \\
	    & = & m_{ai} \left( \exp( i \underline{\xi}\cdot \underline{\tau}/f
\right)^i_j
 \ea

 For the case we are interested here $U$ belongs to $SU_R(2)$ and can be
 expressed in the convenient form,

 \ba
 U & = & \exp[ i \underline{\xi}\cdot \underline{\tau}/f] \\
   & = & \cos||\xi|| + i \widehat{\xi} \cdot \underline{\tau} \sin||\xi||
 \ea

 where $||\xi|| = \sqrt{\xi_1^2 +\xi_2^2 + \xi_3^2}/f$ and
 $\widehat{\xi_i} = \xi_i\slash \sqrt{\xi_1^2 +\xi_2^2 + \xi_3^2}$.

 Recall that we chose the symmetry breaking scheme where
 $SU_R(2)$ was spontaneously broken down to the $U(1)$ generated
 by $\tau_2$.   Thus the Nambu-Goldstone modes point in the
 $\tau_1$ and $\tau_3$ directions.
The most general form of $U$ that we will therefore
 be interested in therefore is of the form,

 \be
 U = \exp[ i (\xi_1 \tau_1 + \xi_3 \tau_3) / f ]
 \ee

 In particular,

 \ba
 U^1_1 & = & \cos||\xi|| + i \widehat{\xi_3} \sin||\xi|| \\
 U^1_2 & = & i  \widehat{\xi_1} \sin||\xi|| \\
       & = & U^2_1 \\
 U^2_2 & = &  \cos||\xi|| - i \widehat{\xi_3} \sin||\xi||
 \ea

 If for concreteness and simplicity we take
  $m_{ai} = m \ \delta_{ai}$ then,

 \be
 \lambda_1^{(2)} = - \frac{m^2}{M} \left[ \cos(2||\xi||) + i \widehat{\xi_3}
\sin(2||\xi||) \right]
 \ee

 and

 \be
 \lambda_2^{(2)} = - \frac{m^2}{M} \left[ \cos(2||\xi||) - i \widehat{\xi_3}
\sin(2||\xi||) \right]
 \ee

 Since the effective potential for the $\xi_i$'s is generated by the
 propogation of neutrinos in the loops the effect of the heavier
 neutrinos is supressed by extra powers of ${m}\slash{M}$.
 Thus,  it is the two lightest neutrinos that contribute most significantly
 to the effective potential for the $\xi$'s.

 Substituting the above eigenvalues back in the expression for
 $-{\cal L}_{\rm{yuk}}$ and recalling that,

 \ba
 \psi_L = \frac{1}{2} ( 1 + \gamma_5) \psi \\
 \psi_R = \frac{1}{2} ( 1 - \gamma_5) \psi
 \ea

 we get that,

 \ba
 -{\cal L}_{\rm{yuk}} & = &  - \frac{m^2}{M}  \overline{\psi}_1 \left[
\cos(2||\xi||) + i \widehat{\xi_3} \sin(2||\xi||) \gamma_5 \right] \psi_1
\nonumber \\
 & &  - \frac{m^2}{M}  \overline{\psi}_2 \left[ \cos(2||\xi||) - i
\widehat{\xi_3} \sin(2||\xi||) \gamma_5 \right] \psi_2
 \ea

\section{The effective potential for our model and its implications}

 We will treat the $\xi_i$'s as classical background fields and not allow them
 to propogate in loops.   Thus,  the effective potential of the  $\xi_i$'s is
 going to be due to the fermion loops.   Thus when we perform the shift
 $ \xi \rightarrow \xi - w $ we need only keep the lowest order term in
 $\xi$.
 (The terms that will contribute to the tadpole diagram have to be of the
 form $ \overline{\psi} \xi \psi $ and can be isolated from the above
 expression for ${\cal L}_{\rm{yuk}}$ by performing the shift
 $ \xi \rightarrow \xi - w $.  )

 When we examine the expression for  ${\cal L}_{\rm{yuk}}$ above we notice
 that on performing the shift $ \xi \rightarrow \xi - w $ we will need
 to expand the terms $|| \xi - w ||$ and
 $\frac{ \xi_3 - w_3}{|| \xi_3 - w_3||}$.
 To linear order in $\xi$,  their expansions are,

 \ba
 || \xi - w || & = & || w || - \frac{\underline{\xi} \cdot \underline{w}}{|| w
||} + O(\xi^2) \\
 \frac{ \xi_3 - w_3}{|| \xi_3 - w_3||} & = & -  \widehat{w_3} + \frac{\xi_3}{||
w ||} - w_3 \frac{\underline{\xi} \cdot \underline{w}}{|| w ||^3} + O(\xi^2)
 \ea

 Using these expansions in  ${\cal L}_{\rm{yuk}}$ and performing
 straightforward trigonometric and algebraic manipulations one can extract
 the following $ \overline{\psi} \xi \psi $ couplings that are all
 relevant for computing the tadpole diagrams that will give us an
 expression for the effective potential.
 The coefficient of the  $ \overline{\psi_1} \xi_3 \psi_1 $ term is:
 $ - \frac{m^2}{M} \left[ y_1 + i \gamma_5 y_2 \right]$,  where,

 \ba
 y_1 & = & \frac{2 w_3 \sin(2 ||w||)}{||w||} \\
 y_2 & = & \frac{w_3^2 \cos(2||w||)}{||w||^2} + \left(\frac{1}{||w||} -
\frac{w_3^2}{||w||^3} \right) \sin(2||w||)
 \ea

 The coefficient of the  $ \overline{\psi_1} \xi_1 \psi_1 $ term is:
 $ - \frac{m^2}{M} \left[ y_3 + i \gamma_5 y_4 \right]$,  where,

 \ba
 y_3 & = & \frac{2 w_1 \sin(2 ||w||)}{||w||} \\
 y_4 & = & \frac{w_3 w_1 \cos(2||w||)}{||w||^2} - \frac{w_3 w_1}{||w||^3}
\sin(2||w||)
 \ea

 The coefficient of the  $ \overline{\psi_2} \xi_1 \psi_2 $ term is
 $ - \frac{m^2}{M} \left[ y_3 - i \gamma_5 y_4 \right]$ and
 the coefficient of the  $ \overline{\psi_2} \xi_3 \psi_2 $ term is
 $ - \frac{m^2}{M} \left[ y_1 - i \gamma_5 y_2 \right]$.

 The tadpole diagram with one external $\xi_1$ leg will receive contributions
 from the loops due to $\psi_1$ and $\psi_3$ and is given by,

 \ba
 \frac{dV_{eff}}{dw_1} & = & \Gamma_{w_1, w_3}^{(1, 0)}  \\
		       & = & - 8 i \int \frac{d^4 k}{(2 \pi)^4} \frac{m^2}{M} \frac{\left[
y_3 A_1 - i y_4 B_1 \right]}{k^2 - (A_1^2 - B_1^2)}
 \ea

 where $y_3, ~ y_4$ were given earlier and $A_1, ~ B_1$ are given by:

 \ba
 A_1 & = & - \frac{m^2}{M} \cos(2 ||w||)  \\
 B_1 & = &  \frac{ i m^2}{M} \left[ \frac{w_3}{||w||} \sin(2||w||) \right]
 \ea

 Similarly the tadpole diagram with one external $\xi_3$ leg will be given
 by,

 \ba
 \frac{dV_{eff}}{dw_3} & = & \Gamma_{w_1, w_3}^{(0, 1)}  \\
		       & = & - 8 i \int \frac{d^4 k}{(2 \pi)^4} \frac{m^2}{M} \frac{\left[
y_1 A_1 - i y_2 B_1 \right]}{k^2 - (A_1^2 - B_1^2)}
 \ea

 The above expressions can be integrated to yield,

 \be
 V_{eff} = 4 i \int  \frac{d^4 k}{(2 \pi)^4} \ln \left( \frac{k^2 -
{\cal{M}}^2(\xi)}{\Lambda^2} + i \epsilon \right)
 \ee

 where,

 \be
 {\cal{M}}^2 = {\frac{m^4}{M^2}} (\cos^2 2 ||\xi|| +
 \widehat{\xi}_3^2 \sin^2 2 ||\xi||)
 \ee

 The above expression can be explicitly evaluated by performing a rotation
 to Euclidean space to yield,

 \ba
 V_{eff}(\xi_1, \xi_3) & = & V_{eff}( {\cal{M}}^2) \\
		      & = & \frac{4}{64 \pi^2} \left[ \frac{\Lambda^4}{2} + 2 \Lambda^2
{\cal{M}}^2 + {\cal{M}}^4 \left(\ln \frac{ {\cal{M}}^2}{\Lambda^2} -
\frac{1}{2} \right) \right]
 \ea

 where as always
$\Lambda$ is the cut-off introduced to regularise the integral.

In order to look for phase transitions in this system we should now add in the
effects of the heat bath.   This should be done with some caution.   The
reason for this is that by the time of decoupling,
the neutrinos have long since
dropped out of thermal equilibrium with the ambient heat bath,  which would
seem
to indicate that the finite temperature formalism  is inapplicable.   However,
we
should recall that despite being out of thermal equilibrium,  neutrinos
preserve
a thermal-like distribution after they decouple (just as the microwave
background photons do),  with a temperature inversely proportional to the scale
factor\cite{Rockybook}.   Thus,  we can still compute the finite tempereature
effective potential
and use it as a diagnostic of critical behavior,  as long as we recall that the
relevant temperature is that of the neutrinos and not that of the heat bath.
Another caveat is that once the temperature is comparable to the mass of the
neutrino,  the distribution function will cease to look thermal.

 The finite temperature effects can be studied by calculating the finite
 temperature corrections to the effective potential which is given by,

 \be
 \Delta V_T = \frac{4 T^4}{2 \pi^2} \int_0^{\infty} dx~ x^2 \ln \left[ 1 +
\exp[- \sqrt{x^2 +\frac{{\cal{M}}^2}{T^2}} \right]
 \ee

 Recall that we parametrized the PNGB's as $\xi_1$ and $\xi_3$
 and defined the quantity
 $\cal{M}$ to be:
 \be
 {\cal{M}}^2 = {\frac{m^4}{M^2}} (\cos^2 2 ||\xi|| +
 \widehat{\xi}_3^2 \sin^2 2 ||\xi||)
 \ee
 Performing the high temperature expansion of the complete potential
 and discarding terms of order $(\m^2)^3\slash T^2$ or higher we get,
 \be
 V_{\rm tot} (\xi_1, \xi_3) = V(\m^2) = (V_0 - \frac{7 \pi^2 T^4}{90}) +
 (m_r^2 + T^2/6)\m^2 + \frac{(\m^2)^2}{8 \pi^2}
 (n - \log \frac{T^2}{\mu^2}),
 \ee
 where $n = 2 \gamma - 1 -2 \log \pi \sim -2.  1303$, $m_r$ is a parameter in
the
 model and $\mu$ is the renormalisation scale.  $\m$ is naturally of the
neutrino
 mass scale in this model.

 A study of the manifold of degenerate vacua of the effective potential at
 different temperatures revealed phase transitions in this model accompanied
 by the formation of topological defects at a temperature of a few times the
 relevant neutrino mass .  Typically at higher temperatures the manifold of
 degenerate vacua consisted of a set of disconnected points whereas at lower
 temperatures the manifold was a set of connected circles.  Thus, domain walls
would
 form at higher temperatures which would evolve into cosmic strings at lower
 temperatures.

An important point is that the potential depends on the $\xi$'s only
through ${\cal M}^2$.   This has the effect of taking a two dimensional problem
and turning it into a one dimensional one,  i.  e.   we need only find the
value
of $\m^2$ that minimizes the effective potential.
A picture of the potential for $m_r^2 = 1$ is given in fig.  1.

We can understand some features of the manifold of degenerate vacuua of this
effective theory if we first uncover some of the symmetries of the potential.
Note that $\m^2$ is left invariant by the transformations:
\ba
||\xi|| \rightarrow ||\xi^\prime || = ||\xi|| + \frac{n\pi}{2} \nonumber \\
\frac{\xi_i}{||\xi||} \rightarrow  \frac{{\xi_i}^\prime}{||\xi^\prime ||}=
\frac{\xi_i}{||\xi||}
\ea
where $n$ is an arbitrary integer.
Furthermore if we define $\hm^2 \equiv \cos^2 2 ||\xi|| +
\widehat{\xi_3}^2 \sin^2 2 ||\xi||$,  we see that $\hm^2$ is restricted to lie
between $0$ and $1$.   It vanishes when $\xi_3 = 0,  \xi_1 = (2k+1)\pi \slash
4$
,  while it attains its maximum value of $1$ when {\em either}
$\xi_1=0$ or $||\xi|| = k \pi\slash 2$ ($k\in {\cal Z}$).   Thus,  if $\hm^2=0$
is
the ground state,  then the vacuum manifold consists of a discrete set of
points,
while if $\hm^2=1$ minimizes the potential,  then this manifold is the set
theoretic union of an
infinite number of circles with discrete radii and the line $\xi_1=0$,  which
intersects all of these circles.   We should note that due to the above
symmetries
some of these vacuua will become identified.   For example,  in the $\hm^2=0$
case
,  the only {\em distinct} vacuua are: $\xi_3 = 0$ and
$\xi_1 = (2 k+ 1)\pi\slash 4$,  with $k = 0,  1,  2,  3$.   For the $\hm^2=1$
situation the distinct vacuua are $\xi_1=0$ or
$||\xi|| = k\pi\slash 2 \ ,  k=0,  1,  2,  3$.

At this point we can define $m_{\nu} = m^2\slash M$ and scale all dimensionful
quantities with this mass.   Thus we define ${\widehat T} = T\slash \mn, \
{\widehat m}_r = m_r\slash \mn$ and we set the renormalization scale $\mu$
equal to $\mn$ (which,  of course,  does not affect any physics).   We then
consider
the dimensionless effective potential
\ba
{\cal V}(\xi_1,  \xi_3) & \equiv & \mn^{-4}
(V_{\rm tot}(\xi_1,  \xi_3)- V_{\rm tot} (\xi_1 = \frac{\pi}{4},  \xi_3 = 0))
\nonumber \\
& = & A(\htemp) \hm^2 + \frac{B(\htemp)}{8 \pi^2} (\hm^2)^2
\ea
where $A(\htemp) = \hmr^2 + \htemp^2\slash 6$,  $B(\htemp) = n - \log
\htemp^2$.

We can now examine the phase structure of ${\cal V}$ as a function of
temperature.   The first fact we can prove is that if $m_r^2 \ge 0$,  then the
ground state is always given by $\hm^2 = 0$.   This is easiest to see by
completing the square in ${\cal V}$ to write it as:
\be
{\cal V}(\xi_1,  \xi_3) = -2 \pi^2 \frac{A(\htemp)^2}{B(\htemp)} +
\frac{B(\htemp)}{8 \pi^2}(\hm^2 + 4 \pi^2 \frac{A(\htemp)}{B(\htemp)})^2
\ee
If ${\widehat T}_0$ is the zero of $B(\htemp)$
($\htempc = \exp (n\slash 2) \sim 0.  34$),
then for $\htemp > \htempc$,  the quantity
$g(\htemp) = - 4 \pi^2 {A(\htemp)}\slash{B(\htemp)}$ is strictly greater than
$2$,  while
for $\htemp < \htempc$ $g(\htemp)$ is negative.   Using these facts,  as well
as
the restriction that $\hm^2$ lies between $0$ and $1$ in plotting the
parabola described by ${\cal V}$ as a function of $\hm^2$ proves our claim.
Thus
domain walls appear when the $\xi_i$'s roll to their minima in the $m_r^2 \ge
0$
case (if initial condidtions allow for this; see below).

The $m_r^2 <0$ case is more interesting.   Set $\alpha = -6 m_r^2\slash
\htempc^2$.
In order that the high temperature approximation
remain valid during our analysis,  we restrict ourselves to values of $\htemp$
satisfying $\htemp >> 1$.

First consider the $\alpha \le 1$ case.   Then for the values of $\htemp$ under
consideration,  $A(\htemp)>0, \ B(\htemp)<0$.   Furthermore,  the location of
the peak
of ${\cal V}$ is at $\hm^2_{\rm max} \sim 0.  77 (x - \alpha)\slash \log x$,
where $x = (\htemp \slash \htempc)^2$.   Now,  $\htemp >> 1$ implies that we
must only
consider values of $x$ larger than $\htempc^{-2} \sim 8.  5$.   In this regime,
$\hm^2_{\rm max}$ is always larger than one,  so that again,  a plot of the
parabola described by ${\cal V}$ as a function of $\hm^2$ leads to the
conclusion that the ground state is at $\hm^2 = 0$,  just as in the $m_r^2\ge
0$
case.

What happens in the $\alpha >1$ case? For $\alpha << \htempc^{-2}$,  we are
essentially back to the situation described in the previous paragraph.
However,
suppose that $\alpha$ is comparable to or larger than $\htempc^{-2}$.   Then it
is
simple to see (using once more the parabola described by ${\cal V}$ as a
function
of $\hm^2$,  see fig.  (2)) that the ground state switches from $\hm^2 = 0$ to
$\hm^2 = 1$ at a critical temperature $\htemp_c$ such that
$-8 \pi^2 A(\htemp_c)\slash B(\htemp_c) = 1$.   This is a second order phase
transition
which takes the domain walls present when $\hm^2 = 0$ was the ground state and
converts them into cosmic strings which are allowed when $\hm^2 = 1$ is the
vacuum state.   This provides a natural solution to the problem of having a
long-lived network of domain walls in theories of late time phase transitions.
We display some snapshots of the phase transition in fig.  (3).

We see then that,  as stated in the introduction,  the vacuum structure of this
class of models is surprisingly rich in terms of the evolution of topological
defects.   We can also find large regions of parameter space for which the
critical temperature of both the phase transition that generates the domain
walls as well as of the one that converts them into strings is
${\cal O}(10 \mn)$.   Thus these models are well suited to being used for late
time phase transition purposes.

Let us now return to considering the observational data that provides the
strongest motivation for a continued investigation of LTPT's.
It has recently been pointed out by M.  Schmidt et al. that comoving quasar
space
densities exhibit a strong peak at redshifts of $2$ to $3$\cite{Scmidt}.  In
their chapter they plot quasar space density as a function of redshift ($z$)
and also separately as a function of cosmic time.  There is an unmistakable
peak in the quasar space density in the region of redshift $2$ to $3$.  The
authors further point out that the observed decline in space density for
space density for $z>3$ is not a result of instrumentational difficulties
in detecting distant quasars.  On the contrary,  the decline in observed quasar
density is a real decline in the number density for $z>3$\cite{v/vmax}.

Thus, Schmidt et al. point out that quasar density peaks sharply at an epoch
of about 2.  3 billion years after the Big Bang.  The full width at half
maximum
of the peak is around 1.  4 billion years.  Their discussion makes the final
point that one needs to understand these time scales in terms of the formation
and evolution of quasars.

What we would like to argue here is that
such a peaked distribution of objects may result from the formation of
topological defects as the universe goes through a phase transition.  Such
topological defects could form the seeds around which quasars light up.

The central power supply of quasars is believed to be gravitational in
origin\cite{REES}.  It is suggested in this chapter that this central power
supply may
be formed as a result of topological defect formation in a phase transition
linked with massive neutrinos.  Such a phase transition would happen at the
right
epoch if one believes the neutrino masses implied by the MSW solution to the
solar neutrino problem.  The production of black holes as a result of
cosmological
phase transitions has been discussed by various authors\cite{KSS,bclref,Widr}.

The idea of topological defects as seeds for structure formation has been
around for some time.  There are many excellent and recent reviews on the
subject\cite{Branden,edcope}.  The idea that topological defects formed
after the decoupling of matter and radiation may play an important role in
structure formation has also been discussed before by Hill, Schramm and
Fry\cite{HillFry}(HSF).

 In the Standard Cosmology the redshift range of $2$ to $3$ corresponds to a
few
 meV in mass scales\cite{Rockybook}.  Thus the critical temperature of the
phase
 transition required to produce quasars is fixed at a few meV.

As we have already pointed out in the context of our particle physics model
for LTPT's we expect a phase transition with a critical temperature
of a few times the relevant light neutrino mass.   Let us therefore now
turn to a discussion of what evidence we have for non-zero masses for
neutrinos and what masses for light neutrinos are implied by these
observations.

 At neutrino detectors around the world,  fewer electron neutrinos are
 received from the sun than predicted by the Standard Solar Model.
 An explanation of the deficiency is offered by the MSW
mechanism\cite{mswtheory}
 which allows the $\nuee$ produced in solar nuclear reactions to change into
 $\numu$.  This phenomenon of neutrino mixing requires massive neutrinos with
the
 masses for the different generations different from each
other\cite{mswtheory}.

 The model we considered earlier was an extremely simple one.  Although it had
2
 families of light neutrinos,  there was only one single light neutrino mass.
As
 such this model was not compatible with the MSW effect.  However it is fairly
 straightforward to modify our original model to make it compatible with the
MSW
 effect as is shown in what follows.

 To ensure that it is
 not possible to choose the  weak interaction eigenstates to coincide with
 the mass eigenstates we must require the 2
 neutrino mass scales to be different.  We can ensure neutrino mixing in our
model
 by demanding that $m_{ai}$ be such that $m_{11} \ne m_{22}$ and
$m_{12}=0=m_{21}
 $.  In this case, the effective potential $V_{\rm tot} (\xi_1,  \xi_3) =
 1/2 (  V( \m _1 ^2) + V( \m _2 ^2) )$ with $ V( \m _i ^2)$ having the same
 functional form as$ V( \m^2) (i=1, 2)$ and $\m _i^2$ given by the following
 expression:
 \be
 {\m _i}^2 = {\frac{m_{ii}^4}{M^2}} (\cos^2 2 ||\xi|| +
 \widehat{\xi}_3^2 \sin^2 2 ||\xi||)
 \ee.

 Further,  if $m_{11}<<m_{22}$  then $V_{\rm tot}(\xi_1,  \xi_3) = V(\m
_2^2)/2$,
 which is exactly half the finite temperature effective potential we discussed
 earlier except the neutrino mass scale is the heavier neutrino mass scale.
Hence,
 the discussion on phase transitions and formation of topological defects we
 carried out earlier goes through exactly except that the critical temperature
is
 determined by the mass scale of the heavier of the 2 neutrinos.

 In the complete picture of neutrino masses\cite{mswtheory},
the neutrinos might
 have a mass hierarchy analogous to those of other fermions.  Further, we
expect
 that the mixing between the first and third generation might be particularly
 small .  Thus we will only consider  $\nuee$-$\numu$ mixing.
 The data seems to imply a central value for the mass of the muon
 neutrino to be a few meV\cite{Bludman}.

 At this point,  let us re-examine the constraint on the domain wall tension,
 $\sigma$ ,  placed by the measurements of the CMBR on small angular scales.
 Recall that the constraint is $\sigma < 0.  5 MeV^3$ .   An estimate of
 $\sigma$ in terms of the quantities $m_\nu$ and $f$ introduced in our model
 can be obtained\cite{Widr}.  (To make contact with the work of L.  Widrow
 cited above please note that his $\lambda m^4 = m^4(\numu)$ and $m = f$
 in our notation.  ) Thus,  the constraint on $\sigma$ then implies that
 $f < 10^{15} GeV$.   Our model is clearly an effective theory with $f$
 being some higher symmetry breaking scale on which it is tough to get
 an experimental handle.   However,  the constraint derived above is in fact
 natural in the context of the see-saw model of neutrino masses embedded
 in Grand Unified Theories as discussed by Mohapatra and Parida\cite{Moha}
 and also by Deshpande, Keith and Pal\cite{Desh}.

 There are a few points on which we'd like to add some further comments.  The
 first has to do with black hole formation as a result of LTPT's.
 The second with the possible role  of a LTPT linked with the mass of
 $\nu_{\tau}$ in the formation of LSS (on scales $\sim 100 Mpc$).
 Finally, a brief comment on the post COBE status of LTPTs is made.

 A number of different groups have studied different mechanisms for black
 hole formation as a result of cosmological phase transitions.  Black hole
 formation as a result of bubble wall collisions has been suggested by
 Hawking et al\cite{hawking}.
 Collapse of trapped false vacuum domains to produce
 black holes
has been studied by Kodama et al.\cite{KSS} and Hsu\cite{Steve}.  The
 collapse of closed domain walls  to form black holes
has been studied by Ipser
 and Sikivie\cite{sikivie}
 and also by Widrow\cite{Widr}.  In fact,  Widrow has suggested that the
 collapse of closed domain walls is  likely to produce black holes
as
 a result of LTPTs.  An estimate of the mass of black holes
produced in LTPTs is
 $M_{BH} \sim 10^9 M_{solar} (\frac{R}{10^3Mpc})^2 (\frac{f}{10^8GeV}) $,
 where $R $ is the radius of the  closed domain wall and we have taken
 $m(\numu)$ to be $3 meV$ to obtain our estimate.

 There are of course two definite uncertainities in getting a number out
 of the above expression.  First, there will obviously be a range of
 possible sizes of closed domain wall bubbles that will be formed in the
 transient period of the phase transition.  A natural upper limit to the
 size of the bubble will be the horizon size at the epoch of the phase
 transitionThus this implies that $R < 10^3 Mpc$.   The horizon size
 at the onset of the phase transition is  smaller.
 Further,  it should
 be noted that sub-horizon sized bubbles which will be formed in the
 transient period of the phase transition may also play an important role.
 Secondly, as already pointed out our model is clearly
 an effective theory with $f$ being some higher symmetry breaking scale
 on which it is tough to get an experimental handle.

 A more detailed analysis of the masses of BH produced in LTPTs as well
 as the number densities of these BH will be
 the subject of a later work.  For now,  the fact that it is possible to
 produce BH which may act as the central engine of quasars is pointed
 out.
 One can check that the mechanism of collapse of domain wall bubbles
 can in fact give roughly the needed number of black holes to power
 quasars.   The Hubble radius at the onset of the phase transition
 at $ z \sim 5 $ ,  is a few hundred $ Mpc $.   One expects a distribution
 of bubble sizes with the horizon as an upper limit to be formed
 during a cosmological phase transition.   In fact,  Turner,
 Weinberg and Widrow\cite{bubs} have carried out a detailed study
 of the distribution of bubble sizes resulting from cosmological
 phase transitions.   They report that typical bubble sizes in a
 successful phase transition range from $0.  01$ to $1$ times the
 Hubble radius at the epoch of the phase transition and depends
 only very weakly on the energy scale of the phase transition.
 Thus,  the most massive black holes formed from the phase
 transition being studied here will have an abundance
 $ \sim 10^{-5}$ to $10^{-6} Mpc^{-3} $ with a higher abundance
 of less massive black holes.   These numbers,  in fact match very well
 with the observational number densities of quasars as discussed
 by Warren and Hewett\cite{WH},  and by Boyle et al. and Irwin et al\cite{BI}.
 The point is that the epoch of the phase transition
 linked with massive neutrinos is about right to explain the observed
 peak in the quasar distribution.  Further,  order of magnitude estimates
 of the masses of black holes produced and their abundances
 are consistent with observations
 within the uncertainities discussed above.  A detailed analysis of the
 efficiency of black hole formation and therefore an accurate number
 for the space density of black holes produced in this model will
 be carried out in a later work.

 The other point that needs to be  commented on is the role of a
 massive $\nu_\tau$ in the scheme of things presented here.  The mass
 of $\nu_\tau$ is less well determined but one would expect an
 earlier phase transition with a $T_c$ of a few times $m(\nu_\tau)$.
 In fact, such a phase transition may well be responsible for the
 LSS seen on the scale of $\sim 100 Mpc$.  There already exists a great deal
 of discussion on this subject.

 However,  black hole formation in this earlier phase transition would be more
 difficult to observe.  In fact, it is the black holes
that are formed in the most recent
 phase transition that will have the greatest observable consequences.

 The anistropy of the CMBR on large angular scales is more closely
 linked to the formation of LSS(on scales $\sim 100$ Mpc).  One can
 relate the power spectrum of density fluctuations to the gravitational
 potential power spectrum responsible for distortions of the CMBR.
 This link is thoroughly discussed
 by Jaffe, Stebbins and Frieman (JSF)\cite{jaffe}.
 Since considerable processing of the power spectrum must take place
 in the process of black hole formation a relationship between
 quasar distribution and the distortions of the CMBR on large angular
 scales is considerably more difficult to establish.
 The viability of LTPTs in the post-COBE era has recently been discussed by
 Schramm and Luo\cite{sclu}.  They point out that LTPTs are still a viable
 model for the formation of LSS.

 To place things
 in perspective one should keep in mind that different mechanisms may play
 important roles in structure formation at different length scales.  This
 point has been emphasised by Carr\cite{carr}who has given a comprehensive
 discusssion on the origin of cosmological density fluctuations.
 In fact,  JSF  also conclude their discussion by pointing out that the final
 power spectrum is most likely due to a combination of primordial and late
 time effects.

\section{Conclusion}

This chapter provides a detailed application of the imaginary time
formalism of finite temperature field theory.
In addition to providing an illustration of many useful
techniques this is also of current cosmological
interest because of a number of recent observations and measurements
that it has the promise of tying up.
Thus, the MSW solution to
the solar neutrino problem seems to imply a
 muon
 neutrino mass of a few meV.  This in turn would lead to a phase transition in
 the PNGB fields associated with massive neutrinos with a critical temperature
 of several meV.  This phase transition happens at the correct epoch in the
 evolution of the universe to provide a possible explanation of the peak in
 quasar space density at redshifts of $2$ to $3$.  This work is clearly
 only a first step in bringing these ideas together.  There are clearly
 some issues which need to be addressed in greater depth and explored in fuller
 detail.   Thus,  a more detailed and accurate analysis of the black hole
 formation efficiency and the masses and
 number densities of BH's needs to be carried
 out.

 To obtain more precise estimates of the black hole number densities and
 masses one needs to make a more detailed analysis of domain formation
 and growth in FRW cosmologies.  Domain formation and growth are among the
 most important non-equilibrium phenomena associated with phase transitions.
 The real time formulation of finite-temperature field theory is the most
 logical and well-suited formalism to study such phenomena.

 Boyanovsky,  Lee and Singh have studied
 domain formation and growth and discussed it at length using
 the real time time formalism\cite{boysinlee}.  This study was carried out in
 Minkowski space.   We need to extend this work to FRW cosmologies.
 In fact,  non-equilibrium phenomena such as particle creation,  entropy
 growth and dissipation have already been studied for FRW cosmologies
 by Boyanovsky, de Vega and Holman\cite{boyvh}.  We now need to examine
 domain formation and growth in this formalism.


 This number density and mass distribution of BH's then needs
 to be compared to the observational data on quasars.  Further,  a more
 detailed
 analysis of the angular dependence of the CMBR anisotropy produced
 needs to be carried out and compared to observations.

Another interesting cosmological implication of the calculation outlined above
is the potential it has of resolving the discrepancy between the recently
measured values of the Hubble constant and the age of the universe.
Increasing improvements in the independent determinations of the
Hubble constant and the age of the universe now seem to indicate that
we need a small non-vanishing cosmological constant to make the two
independent observations consistent with each other.
The cosmological constant can be physically interpreted as due to the
vacuum energy of quantized fields.   To make the cosmological observations
consistent with each other we would need a vacuum energy density,
$ \rho_v \sim (10^{-3} eV)^4 $ today
( in the cosmological units $ \hbar=c=k=1 $ ).
In a recent paper\cite{cosmoc} we have  argued  that such
a vacuum energy density is natural in the context of phase transitions
linked to massive neutrinos.
In fact, the neutrino masses required to solve the cosmological
constant problem
are consistent with those required to solve the solar neutrino problem
by the MSW mechanism.  We will return to a discussion of this subject
in the last chapter.
 My hope is
 that this work will stimulate further thoughts about these issues.

\newpage

\centerline{\bf Figures for Chapter 2:}

\noindent
Figure 1: ${\cal V}$ as a function for $\hm^2$ for temperatures just above,  at
and just below the phase transition temperature.

\vspace{12pt}
\noindent
Figure 2: ${\cal V}(\xi_1,  \xi_3)$ for the temperatures used in fig.  (2).



\newcommand{\x}{\vec{x}}
\newcommand{\k}{\vec{k}}
\newcommand{\vP}{\vec{\Phi}}
\newcommand{\vp}{\vec{\pi}}
\newcommand{\norm}{{\cal{N}}}
\newcommand{\h}{\hbar}
\setlength{\oddsidemargin}{0in}
\setlength{\textwidth}{6.5in}
\setlength{\topmargin}{-0.5in}
\setlength{\textheight}{8.5in}

\chapter{ REAL TIME APPLICATION:  DISSIPATION VIA PARTICLE  PRODUCTION
 IN SCALAR FIELD THEORIES}

\section{Introduction}

In this chapter we will study the dynamics of dissipation in scalar field
theories using the real time formalism. Dissipation and reheating are among
the most important non-equilibrium phenomena that accompany any phase
transition. In cosmology these process are particularly important in the
context of inflation\cite{inflation,reviews}.  Recall that at the end of new or
chaotic
inflation\cite{newchaotic}, defined by when the slow-roll conditions for the
so-called inflaton field\cite{kolbturner,lindebook}, $\phi(t)$, fail to obtain,
the
inflaton begins to oscillate about its true ground state.

The dissipation of the coherent oscillations of the field due to particle
production transfers the enormous vacuum energy of the fields prior to the
phase transition into the energy of the particles. The subsequent
thermalisation of these particles allows the inflationary cosmology to
smoothly merge into the standard hot big bang cosmology. Futher, the
enormous creation of entropy which accompanies the particle production helps
inflation solve the problems of standard cosmology.

In all realistic models of inflation, the inflaton is coupled to
both fermionic and bosonic lighter fields.
As the inflaton oscillates about the
minimum of its potential, it decays to the lighter fields it is coupled to.
This process is usually modeled by taking the evolution equation of the
inflaton $ \phi $ and inserting by hand a term which looks like
$ \Gamma \dot{\phi} $\cite{albrechtetal,abbottwise}.
It is important to state that the $ \Gamma \dot{\phi} $
term is not derivable starting from the fundamental Lagrangian or Hamiltonian
of the system but has to be put in by hand just to model the dissipative
dynamics.
A more careful treatment is both desirable and required and this has been
pointed also by Linde et al and Brandenberger et al\cite{lindebranden}

This chapter will lay out the formalism that allows us to study the detailed
dynamics of dissipation via particle production starting from the exact
microscopic Hamiltonian of the system.

Though we have studied the time evolution of a scalar field coupled to
fermions as well as another scalar field, in this chapter we will study the
dynamics of a self-interacting scalar field. This is a natural first step.
It will allow us to lay out the formalism in the simplest possible setting and
allow us to gain insight into some of the essential physics in this simple
setting.

One of the issues which frequently arises in trying to understand
dissipation from a microscopic perspective is that dissipation appears to be
an irreveresible process whereas the underlying microscopic evolution equation
are time reversible. In this context, it is important to realize the
importance of two important factors that alow us to obtain dissipative
dynamics from time reversal invariant evolution equations. These two
 factors are non-equilibrium initial conditions and coarse graining. Thus,
 even though the evolution equations are time reversal invariant, the
 non-equilibrium initial conditions make the behaviour of the system
irreversible. Further, coarse graining implies that even though total energy
is conserved there is a transfer of energy from the modes of interest into the
remaining modes which now play the role of an environment to the field modes
 or interest are coupled to.

Here is the plan for this chapter.
For a self-interacting scalar field theory we  study the time
evolution of system by having the expectation value of the field away from the
minimum of its potential. This provides the necessary non-equilibrium initial
condition. The expectation value of the field starts oscillating about the
minimum. These oscillations of the expectation value start damping out as the
 fluctuations start building up. We  also study particle production taking
 place in the system. Thus, we arrive at the conclusion that there is a
 precise correlation between the damping out of the oscillations of the zero
 mode with the building up of the fluctuations and particle production.

We  study two different self-interacting scalar field theories. One
possessing a discrete symmetry and another possessing a continuous symmetry.
 For both these systems we  investigate what happens when the symmetry is
unbroken and then when the symmetry is broken. We will notice that the most
dramatic damping takes place in the broken continuous symmetry case because of
 the availability of the Goldstone channel for dissipation.

We  start off by setting up the formalism to study the time evolution. The
time evolution of the system is determined by the quantum Liouville equation.
We are interested in the time evolution of the expectation value of the
field. (This time evolution equation is in fact a functional form of the
Ehrenfest Theorem.) The field can be decomposed into the expectation value and
fluctuations about the expectation value. As a first step we can try to do a
loop expansion. However, we will notice that very soon the one-loop
contribution becomes of the same order as the zeroth order contribution and
 hence one cannot trust the perturbative expansion beyond this time. Thus, one
is driven to consider ways to extract non-perturbative information. We next
use the self-consistent Hartree Approximation to extract information of
non-perturbation dynamics. We will then
introduce the formalism to study particle
 production. The evolution equation we arrive at are coupled
integro-differential equation which we  solve numerically. Plots of the
 time evolution of the expectation value of the field, the growth of
fluctuations and particle production will allow us to gain insight into the
physics of this problems.

\newpage

\section{Non-equilibrium field theory and equations of motion}

The generalization of statistical mechanics techniques to the description of
non-equilibrium processes in quantum field theory has been available for a long
time\cite{schwinger,mahan,keldysh,mills,zhou} but somehow has not yet been
accepted as an integral part of the available tools to study field theory in
extreme environments. We thus begin by presenting  a somewhat pedagogical
introduction to the subject for the non-practitioner.
The discussion that follows is based on understanding gained through
collaborations with Boyanovosky, de Vega, Holman
and Lee\cite{boysinlee,danhec,frw,boyholveg,dissip}

The non-equilibrium description of a system is determined by the time evolution
of the density matrix that describes it.  This time evolution (in the
{Schr\"{o}edinger} picture) is determined by the quantum Liouville equation.
A non-equilibrium situation arises whenever the Hamiltonian
does not commute with the density matrix. Here we allow for
an {\em explicitly} time dependent Hamiltonian, which might be the case
if the system is in an external time dependent background, for example.

Our starting point is the Liouville equation for the density matrix
of the system:
\begin{equation}
i \hbar \frac{\partial \rho (t)}{\partial t} = [H_{\rm evol}, \rho(t)],
\label{liouville}
\end{equation}
where $H_{\rm evol}$ is the Hamiltonian of the system that drives the non-
equilibrium evolution of the system.

Next we use the density matrix to extract the order parameter from the field in
the {Schr\"{o}edinger} picture:

\begin{equation}
\phi(t) = \frac{1}{\Omega}\int \ d^3 x\ {\rm Tr} \left[\rho(t)
\Phi(\vec{x})\right],
\end{equation}
with $\Omega$ the spatial volume
(later taken
to infinity), and $\Phi(\vec{x})$ the field in the {Schr\"{o}edinger} picture.
Using
the Liouville equation together with the Hamiltonian:

\begin{equation}
H = \int  d^3 x \left\{\frac{\Pi^2}{2} + \frac{1}{2} (\nabla \Phi)^2 +
V(\Phi) \right\}
\end{equation}
and the standard equal time commutation relations between a field and its
canonically conjugate momentum, we find the equations:

\begin{eqnarray}
\frac{d \phi(t)}{dt} & = & \frac{1}{\Omega}\int d^3x
  \langle \Pi(\vec{x},t) \rangle
 =\frac{1}{\Omega}\int d^3x  Tr \left[ {\rho}(t) \Pi(\vec{x})\right] = \pi(t)
\label{fidot} \\
\frac{d \pi(t)}{dt}  & = & -\frac{1}{\Omega} \int d^3x \langle
 \frac{\delta V(\Phi)}{\delta \Phi(\vec{x})} \rangle . \label{pidot}
\end{eqnarray}

{}From these equations we can find the equation of motion for the order
parameter
$\phi(t)$:
\begin{equation}
\frac{d^2 \phi(t)}{dt^2}+\frac{1}{\Omega}\int d^3x \langle
\frac{\delta V(\Phi)}{\delta \Phi(\vec{x})} \rangle =0 \label{equationofmotion}
\end{equation}

We expand the field operator as $\Phi(\vec{x}) = \phi(t) +
\psi(\vec{x}, t)$, insert this expansion into equation (\ref{equationofmotion})
and keep only the quadratic terms in the fluctuation field $\psi(\vec{x},t)$.
Doing this yields the equation:

\begin{equation}
\frac{d^2 \phi(t)}{dt^2}+V'(\phi(t))+\frac{V'''(\phi(t))}{2 \Omega}\int d^3x
\langle \psi^2(\vec{x},t)\rangle + \cdots
=0 \label{effequation}
\end{equation}
Here the primes stand for derivatives with respect to $\phi$.

At this point we need to compute and keep track of the fluctuations in the
field. The fluctuations $\langle \psi^2(\vec{x},t)\rangle$ is the two-point
correlation function. In the language of field theory this is the two-point
equal time Green's function. This Green's function can be
evaluated in terms of the homogeneous solutions of the quadratic form for
the fluctuations.For concreteness let us consider a self-interacting scalar
field with the potential,
\begin{equation}
V(\Phi) = \frac{1}{2} m^2 \Phi^2 + \frac{\lambda}{4!} \Phi^4 \label{potential}
\end{equation}

In this case the homogeneous solutions
of the quadratic form for the fluctuations are detemined by
\begin{eqnarray}
\left[\frac{d^2}{dt^2}+\vec{k}^2+m^2+\frac{\lambda}{2}\phi^2(t)
\right]U^{\pm}_k(t) & = & 0 \label{modes} \\
U^{\pm}_k(0)= 1 \; \; ; \; \;
\dot{U}^{\pm}_k(0)  & = & \mp i \omega^0_k
\label{bc}
\end{eqnarray}
with
\begin{equation}
 \omega^0_k          =  \left[\vec{k}^2+m^2+
\frac{\lambda}{2}\phi^2(0)\right]^{\frac{1}{2}}. \label{freqini}
\end{equation}
 The boundary conditions (\ref{bc}) correspond to positive
$U^+$ and negative $U^-$ frequency modes for $t<0$
(the Wronskian of
these solutions is $2i \omega^0_k$).
Notice that $U_k^-(t)=[U_k^+(t)]^*$.
 In terms of these mode functions we obtain the two-point
equal time Green's function to be
\begin{equation}
G(t,t) = \frac{i}{2\omega^0_k}U_k^+(t)U_k^-(t)
\end{equation}
Thus to  order $\hbar$ we find
the following equations
\begin{eqnarray}
\ddot{\phi}(t)+m^2 {\phi}(t)+\frac{\lambda}{6}{\phi}^3(t)+
\frac{\lambda \hbar}{2}{\phi}(t)\int \frac{d^3k}{(2\pi)^3}
\frac{|U^+_k(t)|^2}{2\omega^0_k} & = & 0 \label{oneloopeqn} \\
 \left[\frac{d^2}{dt^2}+\vec{k}^2+m^2+\frac{\lambda}{2}
\phi^2(t)
\right]U^+_k(t)                  & = & 0 \label{modeeqn} \\
U^{+}_k(0)= 1 \; \; ; \; \;
\dot{U}^{+}_k(0)                 & = & -i \omega^0_k
\label{bceqn}
\end{eqnarray}
where we have restored the $\hbar$ to make the quantum
corrections explicit.
This set of equations  clearly shows how the expectation
value (coarse grained variable) ``transfers energy'' to the mode
 functions via a time dependent frequency,
 which then in turn modify the equations of motion
for the expectation value. The equation for the mode functions,
 (\ref{modeeqn}) may be solved in a perturbative expansion in terms
of $\lambda \phi^2(t)$ involving the {\em retarded}
 Green's function.

 Before attempting a numerical solution of the
above equations it is important to understand the renormalization
aspects. For this we need the large $k$ behaviour of the mode
functions which is obtained via a WKB expansion as in references
\cite{boyveg,boyholveg} and to which the reader is referred to
for details. We obtain
\begin{equation}
\int \frac{d^3k}{(2\pi)^3}
\frac{|U^+_k(t)|^2}{2\omega^0_k} = \frac{\Lambda^2}{8\pi}-
\frac{1}{8\pi}\left[m^2+\frac{\lambda}{2}{\phi}^2(t)\right]
\ln\left[\frac{\Lambda}{\kappa}\right]+ \mbox{ finite }
\label{divergences}
\end{equation}
where $\Lambda$ is an ultraviolet cutoff and $\kappa$ an
arbitrary renormalization scale. From the above expression
it is clear how the mass and coupling constant are renormalized.
It proves more convenient to subtract the one-loop contribution at
$t=0$ and absorb a finite renormalization in the mass, finally
 obtaining
the renormalized equation of motion
\begin{eqnarray}
\ddot{\phi}(t)+m_R^2 {\phi}(t)+\frac{\lambda_R}{6}{\phi}^3(t)
+ \frac{\lambda_R \hbar}{8\pi^2}{\phi}(t)\int^{\Lambda}_0 k^2 dk
\frac{\left[|U^+_k(t)|^2-1\right]}{\omega^0_k}
& & \nonumber \\
 + \frac{\lambda^2_R \hbar}{32\pi^2}
{\phi}(t)(\phi^2(t)-\phi^2(0))\ln\left[\Lambda / \kappa\right]
& = &  0 \label{renormeqnofmot}
\end{eqnarray}
In the equations for the mode functions the mass and coupling
may be replaced by the renormalized quantities to this order.
One would be tempted to pursue a numerical solution of these
coupled equations. However doing so would not be consistent,
since these equations were obtained only to order $\hbar$
and a naive numerical solution will produce higher powers of
$\hbar$ that are not justified.

 Within the spirit of the loop
expansion we must be consistent and only keep terms of order
 $\hbar$. First we introduce  dimensionless
variables
\begin{equation}
\eta(t)    =  \sqrt{\frac{\lambda_R}{6 m^2_R}} \phi(\tau)
\; \; ; \; \; \tau        =   m_R t \; \; ; \; \;
q        =  \frac{k}{m_R} \; \; ; \; \;
g = \frac{\lambda_R \hbar}{8\pi^2} \label{dimensionlessquantities}
\end{equation}
and expand the field in terms of
$g$ as
\begin{equation}
\eta(\tau)  =  \eta_{cl}(\tau)+g \eta_1(\tau)+ \cdots \label{fieldexpansion} \\
 \end{equation}
Now the equations of motion consistent up to ${\cal{O}}(\hbar)$
become
\begin{eqnarray}
&&\ddot \eta_{cl}(\tau)+ \eta_{cl}(\tau)+\eta^3_{cl}(\tau)    =  0
\label{classpart} \\
&&\ddot{\eta}_1(\tau)+\eta_1(\tau)+3\eta^2_{cl}(\tau)
                                      \eta_1(\tau) +
\eta_{cl}(\tau)
\int^{\Lambda/m_R}_0 q^2 dq
\frac{\left[|U^+_q(\tau)|^2-1\right]}
{\left[q^2+1+3\eta^2_{cl}(0)\right]^{1/2}}         + \nonumber \\
&&\frac{3}{2}\eta_{cl}(\tau)\left(\eta^2_{cl}(\tau)-\eta^2_{cl}(0)\right)
\ln[\Lambda/m_R]                                     =  0  \label{quancorr}
\end{eqnarray}

The solution to equation (\ref{classpart}) is an elliptic function.
The equations for the mode functions become
\begin{equation}
\left[\frac{d^2}{d\tau^2}+q^2+1+3\eta^2_{cl}(\tau)\right]U^+_q(\tau)
                                                   =  0 \label{dimenmods}
\end{equation}
with the boundary conditions as in eq.(\ref{bceqn}) in terms of the
dimensionless frequencies and
where for simplicity, we have chosen the renormalization scale $\kappa=m_R$.
The chosen
boundary conditions $\eta(0) = \eta_0 \; ; \; \dot{\eta}(0)=0$
can be implemented as
\begin{equation}
\eta_{cl}(0) = \eta_0 \; \; ; \; \; \dot{\eta}_{cl}(0)=0 \; \; ; \; \;
\eta_1(0) =0 \; \; ; \; \; \dot{\eta}_1(0) = 0 \label{eqnsplit}
\end{equation}
In fig.(2) we show $\eta_1(\tau)$ with the above boundary
conditions with $\eta_0=1$ and $g=0.1$. A cutoff $\Lambda /m_R =100$ was
chosen but no cut-off sensitivity was detected by varying the
cutoff by a factor 3. Notice that the amplitude grows as
a function of time. This is  the behavior shown in
fig.(2).

\subsubsection{Failure of perturbation theory to describe dissipation}
This section has been devoted to a perturbative analysis of the
``dissipative aspects'' of the equation of motion for the scalar field.
Perturbation theory has been carried out as an
expansion up to  ${\cal{O}}(\hbar)$, both in the broken and
the unbroken symmetry case.
The potential for the unbroken and broken symmetry cases is given in figures
3 and 4 respectively.
In both cases
 we found that the amplitude of the quantum
corrections {\em grow as a function of time}
and that the long time behavior cannot be captured in perturbation
theory.
This failure of perturbation theory to describe dissipation is
clearly understood from a very elementary but yet illuminating example:
the damped harmonic oscillator. Consider a damped harmonic oscillator
\begin{equation}
\ddot{q}+\Gamma \dot{q}+q=0 \label{dho}
\end{equation}
with $\Gamma \sim {\cal{O}}(\lambda)$ where $\lambda$ is a
small perturbative coupling.
The above equation can be solved exactly to yield the solution,
\begin{equation}
q(t)= e^{-\frac{\Gamma}{2}t}\cos[\omega(\Gamma^2)t] \approx
\cos(t)-\frac{\Gamma}{2}t \cos(t) + {\cal{O}}(\Gamma^2).
\end{equation}
We see that there does
exist a perturbative expansion in $\Gamma$, but in order to find
appreciable damping, we must wait a time $\sim {\cal{O}}(1/\Gamma)$
at which perturbation theory becomes unreliable.

In order to properly describe dissipation and damping one must
resum the perturbative expansion.
Another hint that points to a resummation of the perturbative
series is provided by the set of equations
eqs.(\ref{classpart}-\ref{dimenmods}).
In eq.(\ref{classpart}), the classical solution is a periodic
function of time of constant amplitude, since the classical equation has
a conserved energy. As a consequence, the ``potential'' in the
equation for the modes (eq.(\ref{dimenmods})) is a periodic function of
time with constant amplitude. Thus although the fluctuations
react back on the coarse grained field, only the classically
conserved part of the motion of the coarse grained field enters
in the evolution equations of the mode functions. This is a
result of being consistent with the loop expansion, but clearly
this approximation is  not energy conserving.

As we will point out in the next section, in an energy conserving scheme
the fluctuations and amplitudes will grow up to a maximum value and then
will always remain bounded at all times.

Thus in summary for this section, we draw the conclusion that
perturbation theory is not sufficient (without major ad-hoc assumptions)
to capture the physics of dissipation and damping in real time.
A resummation scheme is needed that effectively sums up the whole
(or partial) perturbative
series in a consistent and/or controlled
manner,
and which provides a reliable estimate for the long-time behavior.
The next section is devoted to the analytical and numerical study
of some of these schemes.

\section{ Non-perturbative schemes I: Hartree approximation}

Motivated by the failure of the loop expansion, we now proceed
to consider the equations of motion in some non-perturbative schemes. First we
study a single scalar model in the time dependent  Hartree approximation. After
this, we study an O(N) scalar theory in the large N limit. This last case
allows us to study the effect of Goldstone bosons on the time evolution of the
order parameter.

In a single scalar model with the interactions described by the
potential of
eq.(\ref{potential}), the Hartree approximation is
implemented as follows. We again decompose the fields as
$\Phi=\phi+\psi$.
The Hartree approximation is implemented by factorising all powers of the
field higher than two and expressing them in terms of products of two point
correlation functions and upto quadratic terms in the field and demanding
that a self-consistency condition be satisfied.
Thus, the Hartree approximation is obtained by assuming the
factorization
\begin{eqnarray}
 \psi^3 ( \vec{x}, t) & \rightarrow & 3 \langle \psi^2 ( \vec{x},t) \rangle
   \psi (\vec{x},t)   \nonumber \\
 \psi^4 ( \vec{x}, t) & \rightarrow & 6 \langle \psi^2 ( \vec{x},t) \rangle
   \psi^2 (\vec{x},t)  -3  \langle \psi^2 ( \vec{x},t) \rangle^2
\end{eqnarray}
Translational invariance shows that
 $\langle \psi^2 ( \vec{x},t) \rangle$ can only be a
function of time.
The expectation value will be determined within a self-consistent
approximation.

In this approximation the resulting Hamiltonian is quadratic, with a
 linear term  in $\psi$:

\be
H_H(t) = \int d^3x \left\{ \frac{\Pi^2}{2}+\frac{(\nabla \psi)^2}{2}
+\psi {\cal{V}}^1(t)+
\frac{{\cal{M}}^2(t)}{2}
\psi^2 \right\} .
 \label{hartreeham1}
\ee

Here $\Pi$ is the canonical momenta
conjugate to $\psi(\x)$  and
${\cal{V}}^1$ is given by,
\be
{\cal{V}}^{(1)}(t) = V'(\phi)+\frac{1}{2}\lambda \phi \langle \psi^2 \rangle
\ee

It is recognized as the derivative of the Hartree
``effective potential''\cite{moshe,camelia} with respect to $\phi$
(it is the derivative of the non-gradient terms of the
effective action\cite{boysinlee,frw,avanvega}).

Also,
\begin{equation}
  {\cal{M}}^2 (t)  = V^{\prime \prime}(\phi) + \frac{\lambda}{2}
    \langle \psi^2 (t) \rangle = m^2 + \frac{\lambda}{2} \phi^2 (t) +
\frac{\lambda}{2}
    \langle \psi^2 (t) \rangle \label{timemass}
\end{equation}
With this Hartree factorization we can use the non-equilibrium
formalism outlined in the previous section to determine the time
evolution.
The resulting equations are obtained to be:
\begin{eqnarray}
   & &\ddot{\phi} + m^2 \phi+ \frac{\lambda}{6} \phi^3 +
 \frac{\lambda}{2} \phi \langle
    \psi^2( t ) \rangle =0 \label{hartreeequation} \\
   & &\langle \psi^2 ( t ) \rangle  =\int \frac{d^3 k}{(2 \pi)^3} \left[
    -i G_{k}(t,t) \right] =
  \int \frac{d^3 k}{( 2 \pi)^3} \frac{\left|U^+_{k} (t)
\right|^2}{2\omega_k(0)}
  \label{qfluctuation}   \\
   & &\left[ \frac{d^2}{d t^2} + \omega_{k}^2 (t) \right]
     U^+_{k} (t)  =0 ~~; ~~~~
  \omega_{k}^2 (t) =\vec{k}^2 + {\cal{M}}^2 (t) \label{modefunction}
\label{hartreeeqns}
\end{eqnarray}
The initial conditions for the mode functions are
\begin{equation}
 U^+_{k} (0) = 1 ~~ ; ~~~~~~~
 {\dot{U}}^{+}_k  (0) = -i \omega_k (0) \label{inmodefunction}
\end{equation}
It is clear that the Hartree approximation makes the
 Lagrangian density quadratic  at the expense of a
 self-consistent condition.
In the time independent case, this approximation sums
up all the ``daisy'' (or ``cactus'') diagrams and leads
to a self-consistent gap equation.

At this stage, we must point out that the Hartree
approximation is uncontrolled in this single scalar theory.
 This approximation does, however, become exact in the
$ N \rightarrow \infty $ limit of the $O(N)$ model which
we will discuss in the next section.

\subsection{ Renormalization}

We now analyze  the renormalization aspects within the Hartree approximation.
To study the renormalization we need to understand the divergences in this
integral
\begin{equation}
 \langle \psi^2 (t) \rangle = \int \frac{d^3 k}{(2 \pi )^3}
   \frac{ \left| U^+_k (t)  \right|^2}{2 \omega_k (0)}
\end{equation}
The divergences  will be determined from the large-k behavior of the mode
functions which obey the differential equations obtained from
(\ref{modefunction}) with
the initial conditions (\ref{inmodefunction}). By a
 WKB-type analysis (see\cite{boyveg,boyholveg} for a detailed description),
 in the $k \rightarrow \infty$ limit, we find
\begin{equation}
  \frac{ \left| U^+_k (t)  \right|^2}{2 \omega_k (0)} =
      \frac{1}{k} -\frac{1}{4 k^3} \left[ m^2 + \frac{\lambda}{2}
    \phi^2 (t)  +
   \frac{\lambda}{2} \langle \psi^{2} (t) \rangle \right] + {\cal{O}}
   ( \frac{1}{k^4} ) + \cdots  \label{renorint}
\end{equation}
Inserting these results in the integral, it is straightforward to find the
divergent terms and we find
\begin{equation}
\int \frac{d^3 k}{(2 \pi )^3}
   \frac{ \left| U^+_k (t)  \right|^2}{2 \omega_k (0)} = \frac{1}{8 \pi^2}
\Lambda^2 - \frac{1}{8 \pi^2} \ln \left(\frac{\Lambda}{\kappa} \right)   \left[
 m^2 + \frac{\lambda}{2}
 \phi^2 (t)   +
   \frac{\lambda}{2} \langle \psi^2 (t) \rangle \right] + finite
\label{divergence}
\end{equation}
where $\Lambda$ is an upper momentum cutoff and $\kappa$ a renormalization
scale.

Now, we are in  position to specify the renormalization
prescription within the Hartree approximation. In this
approximation, there are no interactions, since the
 Lagrangian density  is quadratic. The nonlinearities
are encoded in the self-consistency conditions.
Because of this, there are no counterterms with which to
cancel the divergence and the differential equation for
the mode functions must be finite. Therefore, it leads
 to the following renormalization prescription:
\begin{equation}
  m^2_B + \frac{\lambda_B}{2} \phi^2 (t) +
  \frac{\lambda_B}{2} \langle \psi^2 (t) \rangle_B =
  m^2_R + \frac{\lambda_R}{2} \phi^2 (t) +
 \frac{\lambda_R}{2} \langle \psi^2 (t) \rangle_R
\label{renormalizationprescription}
\end{equation}
where the subscripts $B,R$ refer to the bare and
renormalized quantities respectively and
 $\langle \psi^2 (t) \rangle_B $ is read
from eq.(\ref{divergence}):
\begin{equation}
\langle \psi^2 (t) \rangle_B = \frac{1}{8 \pi^2}
 \Lambda^2 -\frac{1}{8 \pi^2}
 \ln \left( \frac{\Lambda}{\kappa} \right)
\left[   m^2_R + \frac{\lambda_R}{2} \phi^2 (t) +
\frac{\lambda_R}{2} \langle \psi^2 (t) \rangle_R
\right] + \mbox{ finite }
\end{equation}
Using this renormalization prescription eq.(\ref{renormalizationprescription}),
we
obtain
\begin{eqnarray}
   && m^2_B + \frac{\lambda_B}{16 \pi^2} \Lambda^2 = m^2_R \left[ 1+
\frac{\lambda_R}{16 \pi^2} \ln \left( \frac{\Lambda}{\kappa} \right) \right] \\
  && \lambda_B = \frac{\lambda_R}{1- \frac{\lambda_R}{16 \pi^2} \ln \left(
\frac{\Lambda}{\kappa} \right)} \label{renormalizationconditions}
\end{eqnarray}
and
\begin{eqnarray}
 \langle \psi^2 (t) \rangle_R=&&\left[ \frac{1}{1- \frac{\lambda_R}{16 \pi^2}
\ln \left( \frac{\Lambda}{\kappa} \right)}  \right] \times  \nonumber \\
    &&\left\{ \int \frac{d^3 k}{(2 \pi)^3}
 \frac{\left| U^+_k (t) \right|^2}{2 \omega_k (0)}
- \left[  \frac{1}{8 \pi^2} \Lambda^2 -\frac{1}{8 \pi^2}
 \ln \left( \frac{\Lambda}{\kappa} \right) \left(   m^2_R + \frac{\lambda_R}{2}
\phi^2 (t)    \right)  \right] \right\} \nonumber
\end{eqnarray}
It is clear that there is no wavefunction renormalization.
This is a consequence of the approximation invoked. There
 is, in fact, no wavefunction renormalization in either
one-loop or Hartree approximation for a scalar field theory
in three spatial dimensions.

With an eye towards the  numerical analysis,
 it is more convenient to write
\begin{equation}
 \langle \psi^2 (t) \rangle_R =
\left( \langle \psi^2 (t) \rangle_R - \langle \psi^2 (0) \rangle_R \right)+
\langle \psi^2 (0) \rangle_R
\end{equation}
and perform a subtraction at time $t=0$
absorbing $\langle \psi^2 (0) \rangle_R $ into a further
{\em finite} renormalization of the mass term
($ m^2_R +\langle \psi^2 (0) \rangle_R = M^2_R $).

  The renormalized equations that  we will solve finally become
\begin{eqnarray}
   &&\ddot{\phi} + M^2_R \phi + \frac{\lambda_R}{2}
\left[ 1- \left(\frac{2}{3}
\right) \frac{1}{1- \frac{\lambda_R}{16 \pi^2}
\ln \left( \frac{\Lambda}{\kappa}\right)} \right] \phi^3 +
\frac{\lambda_R}{2} \phi \left( \langle \psi^2 (t) \rangle_R - \langle \psi^2
(0) \rangle_R \right)
 =0 \nonumber \\
{}~~~ \\
  &&\left[ \frac{d^2}{dt^2} + k^2 +M^2_R +
\frac{\lambda_R}{2} \phi^2 (t) +
\frac{\lambda_R}{2}\left( \langle \psi^2 (t) \rangle_R - \langle \psi^2 (0)
\rangle_R \right)
\right]U^+_k (t) =0 \label{rhartreeequation}
\end{eqnarray}
and
\begin{eqnarray}
\left( \langle \psi^2 (t) \rangle_R - \langle \psi^2 (0) \rangle_R \right)
 & & = \left[ \frac{1}{1- \frac{\lambda_R}{16 \pi^2} \ln \left(
\frac{\Lambda}{\kappa} \right)}\right] \times \nonumber \\
 & & \left\{ \int \frac{d^3 k}{(2 \pi)^3}
 \frac{\left| U^+_k (t) \right|^2 -1}{2 \omega_k  (0)}
+ \frac{\lambda_R}{16 \pi^2}
 \ln \left( \frac{\Lambda}{\kappa} \right) \left(    \phi^2 (t) -
  \phi^2 (0)  \right)   \right\}  \nonumber
\end{eqnarray}
with the initial conditons for $U^+_k (t) $:
\begin{equation}
 U^{+}_{k} (0) = 1~~ ; ~~~~~~~
 {\dot{U}}^{+}_k  (0) = -i \omega_k (0) \; \; ; \; \;
\omega_k (0) = \sqrt{ k^2 +M^2_R +
\frac{\lambda_R}{2} \phi^2 (0)}
\label{initialumodefunction}
\end{equation}

It is worth noticing  that there is a weak cutoff dependence
 on the renormalized   equations of motion of the order
 parameter and the mode functions.
 This is a consequence of the well known ``triviality'' problem
of the scalar quartic interaction in four space-time dimensions.
This has the consequence  that for a fixed renormalized coupling
the cutoff must be kept fixed and finite. The presence of
the Landau pole prevents taking the limit of the ultraviolet
cutoff to infinity while keeping the renormalized coupling fixed.

 This theory is sensible only as
a low-energy cutoff effective theory. We then must be careful
that for a fixed value of $\lambda_R$, the cutoff must be such
that the theory never crosses the Landau pole.
 Thus from a numerical perspective there will always be
a cutoff sensitivity in the theory. However, for small coupling we expect the
cutoff dependence to be rather weak (this will  be confirmed numerically)
provided the cutoff is  far away from the Landau pole.

\subsection{Particle Production}

Before we engage ourselves in a numerical integration of the
above equations of motion we want to address the issue of
particle production since it is of great importance for the understanding of
dissipative processes.

In what follows, we consider  particle production due to
the time varying effective mass $ {\cal{M}}^2 (t) $ in
eq.(\ref{timemass}) of the quantum field
$ \psi $ for the single scalar model.

In a time dependent background the concept of particle is ambiguous and it
must be defined with respect to some particular state.
Let us consider the Heisenberg fields at $t=0$  written as
\begin{eqnarray}
 && \psi ( \vec{x},0) = \frac{1}{\sqrt{\Omega}} \sum_k \frac{1}{\sqrt{2
\omega_k (0)}} \left( {a}_k (0) + {a}^{\dagger}_{-k} (0) \right)  e^{i \vec{k}
\cdot \vec{x}}  \nonumber \\
 && \Pi_{\psi} ( \vec{x},0) = \frac{-i}{\sqrt{\Omega}} \sum_k
\sqrt{\frac{\omega_k (0)}{2}}
 \left( {a}_k (0) - {a}^{\dagger}_{-k} (0) \right)  e^{i \vec{k} \cdot \vec{x}}
 \label{initialexpansion}
\end{eqnarray}
with $\omega_k (0)$ as in eq.(\ref{modefunction}) and $\Omega$ the spatial
volume.
The  Hamiltonian is diagonalized at $t=0$ by these creation and
destruction operators:
\begin{equation}
 H(0) = \sum_k \omega_k (0) \left[ {a}^{\dagger}_k (0) {a}_k (0) +\frac{1}{2}
\right]
\end{equation}

Thus we define the Hartree-Fock states at $t=0$ as the vacuum annihilated by
$a_k(0)$ together with the tower of excitations obtained by applying
polynomials in $a^{\dagger}_k(0)$ to this vacuum state. The Hartree-Fock vacuum
state at $t=0$ is chosen as the reference state. As time passes, particles (as
defined with respect to this state) will be produced as a result of parametric
amplification\cite{hupa,brantra}. We should mention that our definition differs
from
that of other authors\cite{morikawa,brantra} in that we chose the state at time
$t=0$ rather than using the adiabatic modes (that diagonalize the instantaneous
Hamiltonian).

We define the number density of particles as a function of time as
\begin{equation}
{\cal{N}}(t) = \int \frac{d^3k}{(2\pi)^3}\frac{Tr\left[a^{\dagger}_k(0)a_k(0)
\rho(t)\right]}{Tr \rho(0)}=
\int \frac{d^3k}{(2\pi)^3}\frac{Tr\left[a^{\dagger}_k(t)a_k(t)
\rho(0)\right]}{Tr \rho(0)}
\label{numbdens}
\end{equation}
where by definition
\begin{equation}
a^{\dagger}_k(t) = U^{-1}(t,0)a^{\dagger}_k(0)U(t,0) \; ; \;
a_k(t) = U^{-1}(t,0)a_k(0)U(t,0) \label{heisops}
\end{equation}
are the time-evolved operators in the Heisenberg picture. In terms of these
time
evolved operators, we may write:
\begin{eqnarray}
  \psi ( \vec{x},t) & = & \frac{1}{\sqrt{\Omega}} \sum_k
\frac{1}{\sqrt{2 \omega_k (0)}} \left( {a}_k (t) + {a}^{\dagger}_{-k} (t)
\right)  e^{i \vec{k} \cdot \vec{x}}  \nonumber \\
\Pi_{\psi} ( \vec{x},t)& = & \frac{-i}{\sqrt{\Omega}} \sum_k
\sqrt{\frac{\omega_k (0)}{2}}
 \left( {a}_k (t) - {a}^{\dagger}_{-k} (t) \right)  e^{i \vec{k} \cdot \vec{x}}
 \label{timeexpansion}
\end{eqnarray}

On the other hand, we now expand the Heisenberg fields at time $t$ in the
following orthonormal basis
\begin{eqnarray}
   \psi ( \vec{x},t ) &=& \frac{1}{\sqrt{\Omega}}
\sum_{\vec{k}}\frac{1}{\sqrt{2\omega_k(0)}}
\left( \tilde{a}_k U^+_{k} (t) +
    \tilde{a}^{\dagger}_{-k} U^-_k (t) \right)
 e^{i \vec{k} \cdot \vec{x}} \nonumber \\ \label{psihar}
\Pi_{\psi} ( \vec{x},t ) &=& \frac{1}{\sqrt{\Omega}}
\sum_{\vec{k}}\frac{1}{\sqrt{2\omega_k(0)}}
 \left( \tilde{a}_k \dot{U}^{+}_{k} (t) +
    \tilde{a}^{\dagger}_{-k} \dot{U}^-_k (t) \right)
 e^{i \vec{k} \cdot \vec{x}} \label{pihar}
 \label{orthonormalexpansion}
\end{eqnarray}
where the mode functions $U^+_k(t) \; ; \; U^-_k(t)=(U^+_k(t))^*$ are the
Hartree-Fock mode functions obeying
eqs.(\ref{rhartreeequation}-\ref{initialumodefunction}) together with
the self consistency condition.

Thus $\tilde{a}_k$, $\tilde{a}^{\dagger}_k$ are the   annihilation and creation
operators of Hartree-Fock states, and the Heisenberg field  $\psi (\vec{x},t) $
is a solution of the Heisenberg equations of motion in the Hartree
approximation.  Therefore the  $\tilde{a}_k^{\dagger}$ and $\tilde{a}_k$ do not
depend  on time and are identical to $a_k^{\dagger} (0)$ and $a_k (0)$
respectively ( we can check this  by evaluating the  expansion in
eq.(\ref{orthonormalexpansion}) at $t=0$ together with the  initial conditions
on $U^+_k (t) $ in eq.(\ref{inmodefunction})). Using the Wronskian properties
of the function $U^+_k (t)$, we see that the  $\tilde{a}_k^{\dagger}$ and
$\tilde{a}_k$ satisfy the usual  canonical commutation relations. The reason
for  the choice of the vacuum state at $t=0$ now becomes clear; this is the
initial time at  which the boundary conditions on the modes are determined.The
mode functions  $U^+_k(t)\; ; \;U^-_k(t) $ are then identified with positive
and negative  frequency modes at the initial time.

By comparing the expansion in eq.(\ref{initialexpansion})
 evaluated at time $t$ with that in
eq.(\ref{orthonormalexpansion}), we find that the creation and
annihilation operator at time $t$ can be related to those at
 time $t=0$ via a Bogoliubov transformatiom:
\begin{equation}
   a_k (t) = {\cal{F}}_{+,k} (t) {a}_k (0) +{\cal{F}}_{-,k}
(t) {a}^{\dagger}_{-k} (0)
\end{equation}
The $ {\cal{F}}_{\pm} (t)$ can be read off in terms of the
mode functions  $U^+_k (t)$
\begin{eqnarray}
 \left| {\cal{F}}_{+,k} (t) \right|^2 &=&
\frac{1}{4} \left| U^+_k(t) \right|^2 \left[ 1+
  \frac{ \left| \dot{U}^+_k (t) \right|^2}{\omega^2_k  (0)
\left| U^+_k(t) \right|^2} \right] +\frac{1}{2} \nonumber \\
   \left| {\cal{F}}_{+,k}(t) \right|^2    &-& \left| {\cal{F}}_{-,k} (t)
\right|^2
  =1
   \label{bogoluibov}
\end{eqnarray}

At any time $t$ the expectation value of the number operator for
the quanta of $\psi$ in each k-mode is given by
\begin{equation}
{\cal{N}}_k(t) = \frac{Tr\left[a^{\dagger}_k(t)a_k(t)\rho(0)\right]}{Tr
\rho(0)}
\end{equation}

After some algebra, we find
\begin{equation}
  {\cal{N}}_k(t) = \left( 2  \left| {\cal{F}}_{+,k}(t) \right|^2 -1
 \right)  {\cal{N}}_k (0) + \left( \left| {\cal{F}}_{+,k}(t) \right|^2
-1 \right) \label{numberoperator}
\end{equation}
This result exhibits the contributions from ``spontaneous''
 ( proportional to the initial particle occupations) and
``induced'' ( independent of it) particle production. Since
we are analyzing the zero temperature case with ${\cal{N}}_k(0)=0$ only
the induced contribution results.

\subsection{Numerical Analysis}

\subsubsection{Unbroken Symmetry Case}

In order to perform a numerical analysis it is necessary
to introduce dimensionless quantities and it becomes convenient
to chose the renormalization point $\kappa = M_R$. Thus we define
\begin{eqnarray}
\eta(t) & = & \phi(t) \sqrt{\frac{\lambda_R}{2 M^2_R}}
\; ; \;
q= \frac{k}{M_R} \; ; \; \tau =  M_R t \; ; \; g= \frac{\lambda_R}{8\pi^2}
\nonumber \\
\Sigma(t)
  & = &  \frac{4 \pi^2}{M^2_R}
\left( \langle \psi^2 (t) \rangle_R - \langle \psi^2 (0)
 \rangle_R \right) \label{dimensionless2}
\end{eqnarray}
and finally, the equations of motion become
\begin{eqnarray}
   &&\frac{d^2}{d\tau^2}{\eta} + \eta +
\left[ 1- \left(\frac{2}{3}
\right) \frac{1}{1- \frac{g}{2}
\ln \left( \frac{\Lambda}{M_R}\right)} \right] \eta^3 +
g \eta \Sigma(\tau)
 =0 \nonumber \\
  &&\left[ \frac{d^2}{d\tau^2} + q^2 +1 +
 \eta^2 (\tau) +
g \Sigma(\tau)
\right]U^+_q (\tau) =0 \label{dimrhartreeequation1} \\
  && U^+_q (0)=1 \; ; \; \frac{d}{d \tau}U^+_q (0) = -i
\sqrt{q^2 +1 + \eta^2 (0)} \label{dimbounharcon}
\end{eqnarray}
\begin{eqnarray}
\Sigma(\tau) =  &&
\left[ \frac{1}{1- \frac{g}{2} \ln \left( \frac{\Lambda}{M_R} \right)}\right]
\times \nonumber \\
                       &&\left\{ \int^{\Lambda / M_R}_0 q^2 dq
 \frac{\left| U^+_q (\tau) \right|^2 -1}
{\sqrt{ q^2 +1 + \eta^2 (0)}}
   + \frac{1}{2}
 \ln \left( \frac{\Lambda}{M_R} \right) \left( \eta^2 (\tau)
 -   \eta^2 (0)  \right)   \right\} \label{sigmatau}
\end{eqnarray}

In terms of the dimensionless quantities we obtain the number of
particles within a {\em correlation volume} $N(\tau)={\cal{N}}(t)/M^3_R$
\begin{equation}
N(\tau)=\frac{1}{8\pi^2}\int^{\Lambda / M_R}_0 q^2 dq
\left\{
  \left| U^+_q(\tau) \right|^2+
  \frac{ \left| \dot{U}^+_q (\tau) \right|^2}{\sqrt{ q^2 +1 + \eta^2 (0)}}
  -2 \right\} \label{particleproduction}
\end{equation}

Figures (5.a,b,c) show $\eta(\tau)$, $\Sigma(\tau)$ and  $N(\tau)$ in the
Hartree approximation, for $g=0.1$, $\eta(0)=1.0$ and $\Lambda / M_R =100$; we
did not detect an appreciable cutoff dependence by varying the cutoff between
50 and 200. Clearly there is no appreciable damping in $\eta(\tau)$. In fact it
can be seen that the period of the oscillation is very close to $2\pi$, which
is the period of the classical solution of the linear theory. This is
understood because the coefficient of the cubic term is very small and $g
\Sigma(\tau)$ is negligible. Particle production is also negligible. This
situation should be contrasted with that shown in figures (6.a-c) and (7.a-c)
in which there is  dissipation and damping in the evolution of $\eta(\tau)$ for
$\eta(0)=4,\ 5$ respectively and the same values for $g$ and the cutoff. There
are several noteworthy features that can be deduced from these figures. First
the fluctuations become very large, such that $g\Sigma(\tau)$ becomes
${\cal{O}}(1)$. Second, figures (6.a) and (7.a) clearly show that initially,
channels are open and energy is transferred away from the $q=0$ mode of the
field. Eventually
however, these channels shut off, and the dynamics of the expectation
value settles
into an oscillatory motion. The time scale for the shutting off of the
dissipative behavior decreases as $\eta(0)$ increases; it is about 25 for
$\eta(0)=4$, and about 18 for $\eta(0)=5$. This time scale is correlated with
the time scale in which particles are produced by parametric
amplification and the quantum fluctuations begin to plateau (figures (6.b,c)
and (7.b,c)). Clearly the dissipative mechanism which damps the motion of the
expectation value is particle
production. Furthermore, the long time dynamics for the expectation value
{\em does
not} correspond to exponential damping. In fact, we did not find any
appreciable
damping for $\tau \geq 70$ in these cases. It is illuminating to compare this
situation with that of a smaller coupling depicted in figures (8.a-c) for
$\eta(0)=5$ $g=0.05$. Clearly the time scale for damping is much larger and
there
is still appreciable damping at $\tau \approx 100$. Notice also that
$g\Sigma(\tau) \approx 2$ and that particles are being produced even at long
times and this again correlates with the evidence that the field expectation
value shows
damped motion at long times, clearly showing that the numerical analysis has
not reached the asymptotic regime.

The fundamental question to be raised at this point is: what is the
origin of the damping in the evolution of the field expectation value? Clearly
this is a collisionless process as collisions are not taken into
account in the Hartree approximation (although the one-loop
 diagram that enters in the two particle collision amplitude
with  the two-particle cut is contained in the Hartree approximation and is
responsible for thresholds to particle production).
 The physical mechanism is very similar to that
of Landau damping in the collisionless Vlasov equation for plasmas\cite{landau}
and also found in the study of strong electric fields in
reference\cite{klug}. In the case under consideration, energy
 is transfered from the expectation value
to the quantum fluctuations which back-react on the evolution of
the field expectation value but out of phase.  This phase difference between
the oscillations
of $\eta^2(\tau)$ and those of $\Sigma(\tau)$ can be clearly seen to be
$\pi$ in figures (6.a,b), (7.a,b) since the maxima of $\eta^2(\tau)$ occur at
the same times as
the minima of $\Sigma(\tau)$ and vice versa.

This is an important point learned
from our analysis and that is not {\em a priori} taken into account in the
usual arguments for dissipation via collisions. The process of
thermalization, however, will necessarily involve collisions and
cannot be studied within the schemes addressed in this paper.

\subsubsection{Broken Symmetry}

The broken symmetry case is obtained by writing $M^2_R = -\mu_R^2 < 0$ and
using
the scale $\mu_R$ instead of $M_R$ to define the dimensionless
quantities as in eq.(\ref{dimensionless2}) and the renormalization scale.
The equations of motion in this case become
\begin{eqnarray}
   &&\frac{d^2}{d\tau^2}{\eta} - \eta +
\left[ 1- \left(\frac{2}{3}
\right) \frac{1}{1- \frac{g}{2}
\ln \left( \frac{\Lambda}{\mu_R}\right)} \right] \eta^3 +
g \eta \Sigma(\tau)
 =0 \nonumber \\
  &&\left[ \frac{d^2}{d\tau^2} + q^2 -1 +
 \eta^2 (\tau) +
g \Sigma(\tau)
\right]U^+_q (\tau) =0 \label{dimbrorhartreeequation} \\
  && U^+_q (0)=1 \; ; \; \frac{d}{d \tau}U^+_q (0) = -i
\sqrt{q^2 +1 + \eta^2 (0)} \label{dimbouncon}
\end{eqnarray}
with $\Sigma(\tau)$ given in eq.(\ref{sigmatau}) but with $M_R$
replaced by $\mu_R$. The broken symmetry case is more subtle
because of the possibility of unstable modes for  initial
conditions in which $\eta(0) <<1$. We have kept the boundary conditions
of eq.(\ref{dimbouncon}) the same as in eq.(\ref{dimbounharcon}). This
corresponds to preparing an initial state as a Gaussian state
centered at $\eta(0)$ with real and positive covariance (width)
and letting it evolve for $t>0$ in the
broken symmetry potential\cite{boyveg,dcc}.
The number of particles produced within a correlation volume (now $\mu_R^{-3}$)
is
given by eq.(\ref{particleproduction}) with $M_R$ replaced by $\mu_R$.

Figures (9.a-c) depict the dynamics for a broken symmetry case
in which $\eta(0)=10^{-5}$ i.e. very close to the top of the potential
hill. Notice that as the field expectation value rolls down the hill the
unstable modes
make the fluctuation grow dramatically until about $\tau \approx 50$ at
which point $g\Sigma(\tau) \approx 1$. At this point, the unstable
growth of fluctuations shuts off and the field begins damped oscillatory
motion around a mean value of about $\approx 1.2$.
This point is a minimum of the effective action.
 The damping of these
oscillations is very similar to the damping around the origin in the
unbroken case. Most of the particle production and the largest quantum
fluctuations occur when the field expectation value
is  rolling down the  region for which there are unstable
frequencies for the mode functions
 (see eq.(\ref{dimbrorhartreeequation})).
This behavior is similar to that found previously by some of these
authors\cite{boyveg}.

\section{ Non-perturbative schemes II: Large N limit in the $ O(N)$ Model}

Although the Hartree approximation offers a non-perturbative resummation of
select terms, it is not  a consistent approximation because there is no
{\it a priori} small parameter that defines the approximation. Furthermore
we want to study the effects of dissipation by Goldstone bosons in a non
perturbative but controlled expansion.

In this section, we consider the $O(N)$ model in the large N limit.
The large N limit has been used in studies of non-equilibrium
dynamics\cite{mottola1,mottola2,mottola3,dcc} and it provides a very powerful
tool for studying non-equilibrium dynamics non-perturbatively in a consistent
manner.
The Lagrangian density is
the following:
\begin{eqnarray}
 {\cal{L}} &=& \frac{1}{2} \partial_{\mu} \vec{\phi} \cdot \partial^{\mu}
    \vec{\phi} - V(\sigma, \vec{\pi} ) \nonumber  \\
   V(\sigma, \vec{\pi} ) &=& \frac{1}{2} m^2 \vec{\phi} \cdot \vec{\phi} +
   \frac{ \lambda}{8N} (  \vec{\phi} \cdot \vec{\phi} )^2
\label{onLagrangian}
\end{eqnarray}
for $\lambda$ fixed in the large $N$ limit. Here $\vec{\phi}$ is an $O(N)$
vector,  $\vec{\phi} = (\sigma, \vec{\pi} )$ and $\vec{\pi}$ represents the
$N-1$ pions. In what follows, we will consider two different cases of the
potential $  V(\sigma, \vec{\pi} ) $, with ($m^2 < 0$) or without ($m^2 > 0$)
symmetry breaking.

 Our first order of business is to identify the correct order
parameter for the phase transition and then to obtain its equation of
 motion.
Let us define the fluctuation field operator $ \chi(\x,t)$ as

\be
\sigma = \phi_0(t) + \chi(\x,t),
\ee
with $\phi_0(t)$ a c-number field defined by:

\ba
\phi_0(t) & = & \frac{1}{\Omega} \int d^3 x \langle \sigma(\x) \rangle
 \nonumber \\
        & = & \frac{1}{\Omega} \int d^3 x \frac{{\rm{Tr}}(\rho(t)
\sigma(\x))}{{\rm{Tr}}\rho(t)} .
\ea

We can use the Liouville equation as before to arrive at the following
evolution equation for the order parameter
$\phi_0(t)$:

\be
\ddot{\phi_0}(t) + \frac{1}{\Omega}\int d^3x\
\langle
\frac{\delta V(\sigma, \vp)}{\delta \sigma(\vec{x})}
\rangle = 0.
\ee

To proceed further we have to determine the density matrix. Since the
Liouville equation is first order in time we
need only specify $\rho(t=0)$.
At this stage
we could proceed to a perturbative description of the
dynamics (in a loop expansion).

However, as we learned previously in a
similar situation\cite{boysinlee,danhec} and as we have argued in this
chapter,
the non-equilibrium
dynamics of the phase transition cannot be studied within perturbation
theory.

In the presence of a vacuum expectation value, the Hartree factorization is
somewhat subtle. We will make a series of {\it assumptions} that we feel are
quite reasonable but which, of course, may fail to hold under some
circumstances and for which we do not have an {\it a priori} justification.
These are the following: i) no cross correlations between the pions and the
sigma field, and ii) that the two point correlation functions of the pions are
diagonal in isospin space, where by isospin we now refer to the unbroken
$O(N)$
symmetry under which the pions transform as a triplet. These assumptions
lead to the following Hartree factorization of the non-linear terms in the
Hamiltonian:

\begin{eqnarray}
   \chi^4  & \rightarrow & 6 \langle \chi^2 \rangle \chi^2 + constant \\
   \chi^3  & \rightarrow & 3 \langle \chi^2 \rangle \chi \\
   \left( \vec{\pi} \cdot \vec{\pi} \right)^2 & \rightarrow & \left( 2+
    \frac{4}{N-1} \right) \langle \vec{\pi}^2 \rangle \vec{\pi}^2 + constant \\
   \vec{\pi}^2 \chi^2 & \rightarrow & \langle \vec{\pi}^2  \rangle \chi^2
  +\vec{\pi}^2 \langle \chi^2 \rangle \\
   \vec{\pi}^2 \chi & \rightarrow & \langle \vec{\pi}^2  \rangle \chi
\end{eqnarray}
where by ``constant'' we mean the operator independent expectation values of
the composite operators which will not enter into the time evolution equation
of the order parameter.

In this approximation the resulting Hamiltonian is quadratic, with a
 linear term  in $\chi$:

\be
H_H(t) = \int d^3x \left\{ \frac{\Pi^2_{\chi}}{2}+
\frac{\vec{\Pi}^2_{\pi}}{2}+\frac{(\nabla \chi)^2}{2}+
\frac{(\nabla \vec{\pi})^2}{2}+\chi {\cal{V}}^1(t)+
\frac{{\cal{M}}^2_{\chi}(t)}{2}
\chi^2+\frac{{\cal{M}}^2_{\pi}(t)}{2}\vec{\pi}^2 \right\} .
 \label{hartreeham}
\ee

Here $\Pi_{\chi}, \ \vec{\Pi}_{\pi}$ are the canonical momenta
conjugate to $\chi(\x), \ \vec{\pi}(\x)$ respectively and
${\cal{V}}^1$ is recognized as the derivative of the Hartree
``effective potential''\cite{moshe,camelia} with respect to $\phi_0$
(it is the derivative of the non-gradient terms of the
effective action\cite{boysinlee,frw,avanvega}).

To obtain a large N limit, we define
 \begin{eqnarray}
 \langle
\vec{\pi}^2 \rangle &=& N \langle
\psi^2 \rangle  \label{largeN} \\
\phi_0(t)         &=& \phi(t) \sqrt{N} \label{filargeN}
\end{eqnarray}
with
\begin{equation}
  \langle
\psi^2 \rangle \approx {\cal{O}} (1) \; ; \;  \langle
\chi^2 \rangle \approx {\cal{O}} (1) \; ; \;  \phi \approx  {\cal{O}} (1).
\label{order1}
\end{equation}
 We will approximate further by neglecting the $ {\cal{O}} (\frac{1}{N})$
 terms in the formal large $N$ limit. We now obtain
\begin{eqnarray}
   V^{'} (\phi(t),t)         &=&{\sqrt{N}}
  \phi (t) \left[ m^2 +\frac{\lambda}{2} \phi^2 (t) + \frac{\lambda}{2} \langle
\psi^2 (t) \rangle \right]
\label{vprime} \\
  {\cal{M}}^2_{\vec{\pi}}(t) &=& m^2 + \frac{\lambda}{2} \phi^2 (t) +
  \frac{\lambda}{2}\langle \psi^2 (t) \rangle  \\
  {\cal{M}}^2_{\chi}(t)      &=& m^2 + \frac{3 \lambda}{2} \phi^2 (t) +
  \frac{\lambda}{2} \langle \psi^2 (t) \rangle
\end{eqnarray}
We obtain the following set of equations:
\begin{eqnarray}
&& \ddot{\phi}(t) +  \phi (t) \left[ m^2 +\frac{\lambda}{2} \phi^2 (t) +
\frac{\lambda}{2} \langle \psi^2 (t) \rangle  \right]
   =0 \nonumber \\
&&  \langle \psi^2 ( t ) \rangle =
   \int \frac{d^3 k}{( 2 \pi)^3} \frac{\left| U^{+}_k (t) \right|^2
}{2\omega_{\vec{\pi} k}^2 (0)}
    \label{pionmodefunction}  \\
&& \left[ \frac{d^2}{d t^2} + \omega_{\vec{\pi} k}^2 (t) \right]
     U^{+}_{k} (t)  =0 \; ; \;
  \omega_{\vec{\pi} k}^2 (t) = \vec{k}^2 + {\cal{M}}_{\vec{\pi}}^2 (t)
\end{eqnarray}
The initial conditions for the mode functions $U^+_k (t) $ are
\begin{equation}
 U^{+}_{k} (0) = 1 ~~ ; ~~~~~~~
 {\dot{U}}^+_k  (0) = -i {\omega_{\vec{\pi} k} (0)}
\label{oninmodefunctions}
\end{equation}
Since in this approximation, the dynamics for the $\vec{\pi}$ and $\chi$
fields decouples, and the dynamics of $\chi$ does not influence that of
$\phi$ or the mode functions and $\langle \psi^2 \rangle$
 we will only concentrate on the solution for the $\vec{\pi}$ fields.

\subsection{Renormalization}

The renormalization procedure is exactly the same as that for the Hartree
case in the previous section
(see section 3.3.1).
We carry out the same renormalization prescription and subtraction at $t=0$ as
in the last section.
Thus we find
the following equations of motion

\begin{eqnarray}
   &&\ddot{\phi} + M^2_R \phi + \frac{\lambda_R}{2}  \phi^3
+\frac{\lambda_R}{2} \phi \left(  \langle \psi^2 (t) \rangle_R - \langle \psi^2
(0) \rangle_R  \right) =0  \nonumber\\ \label{eqofmotlarN}
   &&\left[ \frac{d^2}{dt^2} + k^2 +M^2_R + \frac{\lambda_R}{2} \phi^2 (t) +
\frac{\lambda_R}{2} \left( \langle \psi^2 (t) \rangle_R - \langle \psi^2 (0)
\rangle_R \right)  \right] U^+_{k} (t) =0 \label{onrehartreeequation}
\end{eqnarray}
with the initial conditions given by eq.(\ref{oninmodefunctions})
and
with the subtracted expectation value given in section 3.3.1.

In contrast with the Hartree equations in the previous section,
the cutoff dependence in the term proportional to $\phi^3$
in eq.(\ref{eqofmotlarN})
has disappeared. This is a consequence of the large N limit and the Ward
identities, which are now obvious at the level of the renormalized equations of
motion. There is still a very weak cutoff dependence
because of the triviality issue which is not
relieved in the large N limit, but again, this theory only makes sense as a
low-energy cutoff theory.

\subsection{Numerical Analysis}

\subsubsection{Unbroken symmetry}
To solve the evolution equations numerically, we
now introduce dimensionless quantities as in eq.(\ref{dimensionless2})
obtaining the following dimensionless equations:
\begin{eqnarray}
   &&\frac{d^2}{d\tau^2}{\eta} + \eta +   \eta^3 +
g \eta \Sigma(\tau)
 =0 \nonumber \\
  &&\left[ \frac{d^2}{d\tau^2} + q^2 +1 +
 \eta^2 (\tau) +
g \Sigma(\tau)
\right]U^+_q (\tau) =0 \label{dimrhartreeequation} \\
  && U^+_q (0)=1 \; ; \; \frac{d}{d \tau}U^+_q (0) = -i
\sqrt{q^2 +1 + \eta^2 (0)} \label{dimbocon}
\end{eqnarray}
\begin{eqnarray}
\Sigma(\tau) =  &&
\left[ \frac{1}{1- \frac{g}{2} \ln \left( \frac{\Lambda}{M_R} \right)}\right]
\times \nonumber \\
    &&\left\{ \int^{\Lambda / M_R}_0 q^2 dq
 \frac{\left[\left| U^+_q (\tau) \right|^2 -1\right]}
{\sqrt{ q^2 +1 + \eta^2 (0)}}
   + \frac{1}{2}
 \ln \left( \frac{\Lambda}{M_R} \right) \left( \eta^2 (\tau) -
 \eta^2 (0)  \right)   \right\} \label{sigmun}
\end{eqnarray}

For particle production in the $O(N)$ model, the final
expression of the  expectation value of the number operator
 for each pion field in terms of dimensionless
 quantities is the same as in eq.(\ref{particleproduction}),
but the mode function $U^+_q (t)$ obeys the differential
equation in eq.(\ref{dimrhartreeequation1}) together with the
self consistent condition.

Figures (10.a-c), (11.a-c) and (12.a-c) show $\eta(\tau)\; ; \; \Sigma(\tau)\;
; \;
N(\tau)$ for $\eta(0)=1;2$  $g=0.1;0.3$ and $\Lambda / M_R=100$
(although again we did not find cutoff sensitivity). The dynamics is very
similar to that of the single scalar field in the Hartree approximation,
which is not
surprising, since the equations are very similar (save for the coefficient
of the cubic term in the equation for the field expectation value).  Thus the
analysis presented
previously for the Hartree approximation remains valid in this case.

\subsubsection{Broken symmetry}
The broken symmetry case corresponds to choosing $M^2_R = -\mu^2_R <0$. As in
the
case of the Hartree approximation, we now choose $\mu_R$ as the scale to
define dimensionless variables and renormalization scale. The equations of
motion
for the field expectation value and the mode functions now become
\begin{eqnarray}
   &&\frac{d^2}{d\tau^2}{\eta} - \eta +   \eta^3 +
g \eta \Sigma(\tau)
 =0 \nonumber \\
  &&\left[ \frac{d^2}{d\tau^2} + q^2 -1 +
 \eta^2 (\tau) +
g \Sigma(\tau)
\right]U^+_q (\tau) =0 \label{largenmod}
\end{eqnarray}
with $\Sigma(\tau)$ given by eq.(\ref{sigmun}).
As in the  Hartree case, there is a subtlety with the boundary conditions for
the mode functions because the presence of  the instabilities at $\tau=0$ for
the band of wavevectors $0 \leq q^2 <1$, for $\eta^2(0)<1$. Following the
discussion in the
Hartree case (broken symmetry) we chose the initial conditions for the mode
functions as
\begin{equation}
 U^+_q (0)=1 \; ; \; \frac{d}{d \tau}U^+_q (0) = -i
\sqrt{q^2 +1 + \eta^2 (0)}
\end{equation}
These boundary conditions correspond to preparing a gaussian state centered
at $\eta(0)$ at $\tau=0$ and letting this initial state evolve in time in the
``broken symmetry potential''\cite{boyveg}.

Figures (13.a-c) show the dynamics of $\eta(\tau)$, $\Sigma(\tau)$
 and $N(\tau)$  for $\eta(0)=0.5$ $g=0.1$
and cutoff $\Lambda / \mu_R=100$. Strong damping behavior is evident, and the
time scale of damping is correlated with the time scale for growth of
$\Sigma(\tau)$
and $N(\tau)$. The asymptotic value of $\eta(\tau)$ and $\Sigma(\tau)$,
$\eta(\infty)$ and $\Sigma(\infty)$ respectively satisfy
\begin{equation}
-1 +  \eta^2 (\infty) +
g \Sigma(\infty)=0 \label{goldsto}
\end{equation}
as we confirmed numerically. Thus the mode functions are ``massless''
describing
Goldstone bosons. Notice that this value also corresponds to $V'(\phi)=0$
in eq.(\ref{vprime}). An equilibrium self-consistent solution of the
equations of motion for the field expectation value and the fluctuations is
reached for
$\tau= \infty$.

Figure (13.d) shows the number of particles produced per
correlation volume as a function of (dimensionless) wavevector $q$
 at $\tau=200$.
We see  that it is strongly peaked at $q=0$ clearly showing that the particle
production mechanism is most efficient for long-wavelength Goldstone bosons.
At
long times, the contributions from $q \neq 0$ becomes small.
Figures (14.a-c) show the evolution of $\eta(\tau)$, $\Sigma(\tau)$ and
$N(\tau)$
for $\eta(0)=10^{-4}$, $g=10^{-7}$. The initial value of $\eta$ is very
close to the ``top'' of the potential. Due to the small coupling and the
small initial value of $\eta(0)$, the unstable modes (those for which
$q^2 < 1$) grow for a long time making the fluctuations very large. However
the fluctuation term $\Sigma(\tau)$ is multiplied by a very small coupling
and it has to grow for a long time to overcome the instabilities. During
this time the field expectation value rolls down the potential hill, following
a trajectory
very close to the classical one.
  The
classical turning point of the trajectory beginning very near  the top
of the potential hill, is close to $\eta_{tp}=\sqrt{2}$.  Notice that
$\eta(\tau)$ exhibits a turning point (maximum)
 at $\eta \approx 0.45$. Thus the turning point of the effective
evolution equations is much
closer to the origin. This phenomenon shows that the effective (non-local)
 potential is shallower than the classical potential, with the minimum moving
closer to the origin as a function of time.

If the energy for the field expectation value was absolutely conserved,
the expectation value of the scalar field would bounce back to the initial
point and oscillate between the
two classical turning points. However, because
the fluctuations are growing and energy is transferred to them from the
$q=0$ mode, $\eta$ is slowed down as it bounces back, and tends to settle at a
value very close to the origin (asymptotically about 0.015). Figure (14.b)
shows that the fluctuations grow initially and stabilize at a value for which
$g\Sigma(\infty) \approx 1$. The period of explosive growth of the fluctuations
is correlated with the strong oscillations at the maximum of $\eta(\tau)$.
This is the time when the fluctuations begin to effectively absorb the energy
transferred by the field expectation value and when the damping mechanism
begins to work.
Again the asymptotic solution is such that $-1 +  \eta^2 (\infty) +
g \Sigma(\infty)=0 $, and the particles produced are indeed Goldstone bosons
but the value of the scalar field in the broken symmetry  minimum is very
small
(classically it would be $\eta_{min}=1$, yet dynamically, the field
settles at a value $\eta(\infty) \approx 0.015$!! for $g=10^{-7}$).
Figure (14.c) shows copious particle production, and the asymptotic final state
is a highly excited state with a large number (${\cal{O}}(10^5)$)
 of Goldstone bosons per correlation volume.

The conclusion that we reach from the numerical analysis is that Goldstone
bosons are {\it extremely effective} for dissipation and damping. Most of the
initial potential energy of the field is converted into particles (Goldstone
bosons) and the field expectation value comes to rest at long times at a
position very close
to the origin.

Notice that the difference with the situation depicted in figures (13.a-c)
is in the initial conditions and the strength of the coupling. In the case
of stronger coupling, the fluctuations grow only for a short time because
$g\Sigma(\tau)$ becomes ${\cal{O}}(1)$ in short time, dissipation begins to
act rather rapidly and the expectation value rolls down only for a short span
and
comes to rest at a minimum of the effective action,
having transferred all of its potential energy difference to produce Goldstone
bosons.

Figures (15.a-c) show a very dramatic picture. In this case
$\eta(0)=10^{-4}$,
$g = 10^{-12}$. Now the the field begins very close to the top of the
potential hill. This initial condition corresponds to a ``slow-roll''
scenario. The fluctuations must grow for a long time before $g\Sigma(\tau)$
becomes ${\cal{O}}(1)$, during which the field expectation value evolves
classically
reaching the classical turning point at $\eta_{tp} = \sqrt{2}$,
and then bouncing back. But
by the time it gets near the origin again, the fluctuations have grown
dramatically absorbing most of the energy of the field expectation value and
completely
damping its motion. In this case {\em almost all} the initial
potential energy has been converted into particles.
This is a remarkable result. The conclusion of this analysis is that the strong
dissipation by Goldstone bosons dramatically changes the dynamics of the phase
transition. For slow-roll initial conditions the scalar field relaxes to a
final value which is very close to the origin. This is the minimum of the
effective
action, rather than the minimum of the tree-level effective potential. Thus
dissipative effects by Goldstone bosons introduce a very strong dynamical
correction
of the effective action leading to a very shallow
effective potential (the effective action for constant field configuration).
The condition for this situation to happen is that the period of the classical
trajectory is of the same order of magnitude as the time scale of growth for
the fluctuations.

The weak coupling estimate for this dynamical non-equilibrium time scale
is $\tau_s \approx \ln(1/g)/2$, which is
obtained by requiring that $g\Sigma(\tau) \approx 1$. For weak coupling the
mode functions grow as $U^+_q(\tau) \approx e^{\tau}$ for $q^2 <1$, and
$\Sigma(\tau) \approx e^{2\tau}$. The numerical analysis confirms this time
scale for weak coupling. For weakly coupled theories, this non-equilibrium
 time
scale is much larger than the static correlation length (in units in which
$c=1$) and the only relevant time scale for the dynamics.

Clearly the outcome of the non-equilibrium evolution will depend on the initial
conditions of the field expectation value.

Our results pose a fundamental question: how is it possible to reconcile
damping and dissipative behavior, as found in this work with time reversal
invariance?

In fact we see no contradiction for the following reason: the dynamics is
completely determined by the set of equations of motion for the field
expectation value and
the mode functions for the fluctuations described above. These equations are
solved by providing non equilibrium initial conditions on the field expectation
value, its derivative and the
mode functions and their derivatives at the initial time $t=0$. The problem is
then evolved in time by solving the coupled {\em second order} differential
equations. We emphasize the fact that the  equations are second order in time
because these are time reversal invariant. Now consider evolving this set of
equations up to a positive time $t_0$, at which we stop the integration and
find the value of the field expectation value, its derivative, the value of
{\em all} the
mode functions and their derivatives. Because this is a system of
differential equations which is second order in time, we can take these values
at $t_0$ as  initial
conditions at this time and evolve {\em backwards} in time reaching the initial
values  at $t=0$. Notice that doing this involves beginning at a time $t_0$ in
an excited state with (generally) a large number of particles. The conditions
at this particular time are such that the energy stored in this excited state
is focused in the back reaction to the field expectation value that acquires
this energy and
whose amplitude will begin to grow.

It is at this point where one recognizes the fundamental necessity of the
``in-in'' formalism in which the equations of motion are real and causal.

This concludes our discussion of the deatiled numerical analysis of the
dissipation in self-interacting scalar field theories. Let us now turn
to the conclusion of this chapter where we will recapitulate what we have
achieved, place our work in perspective and discuss the future directions
that need to be explored.

\newpage

\section{Conclusion}

In this chapter we have used the real time formalism to study one
of the most important time dependent and non-equilibrium
phenomena that follows every phase transition namely dissipation
and reheating.
Phase transitions play an important role in cosmology. The universe we live in
is an interesting place because of out of equilibrium
phenomena.\cite{kolbturner}
Thus, it is important to systematically analyze the non-equilibrium
phenomena associated with phase transitions.

The real time formalism of finite temperaure field theory is the natural tool
to use in understanding the non-equilibrium phenomena associated with phase
transitions.
We have previously used this formalism to study domain formation and
growth.\cite{boysinlee}
The dynamics of dissipation and reheating that occur during a phase transition
are among the most interesting non-equilibrium phenomena. In  particular,
investigating this in depth is important for understanding the consequences
of inflation\cite{lindebranden}
and other cosmological phase transitions.This chapter was devoted
to understanding the damping of coherent field oscillations due to the
phenomenon of particle production which is the microscopic mechanism for
quantum dissipation.We  studied the simplest case of self-interacting
scalar filed theories to enable us to concentrate on the basics. Here, we
considered a system with a discrete as well as another system with a continuous
internal symmetry in both the unbroken and broken symmetry case. Our results
show that damping of oscillations is most efficient for the broken continuous
internal symmetry case because of the availibility of the goldstone channel
for dissipation.

Our investigations of non-equilibrium phenomena in phase
transitions\cite{boysinlee}
have
consistently shown the neccesity of accounting for non-perturbative
effects when studying the long-time behaviour of the system.\cite{dissip}
If one restricts oneself to the one-loop evolution equations one quickly
sees that the first order effects grow to become of the same order as the
zeroth order effects thus invalidating the use of perturbation theory for
studying late time effects.
To extract the non-perturbative results we have used
the self-consistent Hartree approximation.

The Hartree approximation is implemented by factorising all powers of the
field higher than two and expressing them in terms of products of two point
correlation functions and upto quadratic terms in the field and demanding
that a self-consistency condition be satisfied. Once this is done one still
needs to take care of divergences in the integrals that appear in the
evolution equations. This is done by renormalisation using the WKB technique.

We must point out that the Hartree approximation is uncontrolled in the
single scalar theory. This approximation becomes
exact in the $ N \rightarrow \infty $ limit of an $O(N)$ model which we have
discussed.
In our investigations, we also take the large $N$ limit that is
known as a systematic and controlled  expansion in powers
of $  \frac{1}{N}$ for which the subleading corrections can be studied.

By solving,plotting and studying the solutions to the evolution equations
we were
able to understand many interesting features of interacting systems. In
particular, we  gained an insight into the microscopic
mechanisms of dissipation for interacting systems.
The solution to the coupled integro-differential equations describing the
time evolution for these interacting system was obtained numerically.

We have  discussed two self-consistent non-perturbative approximation schemes
in this paper, The Hartree approximation and the large N approximation for the
O(N) model. Of these the large N approximation has the advantage that it is a
controlled approximation to which the corrections can be calculated in a
systematic manner.In fact, the solutions to the evolution equations obtained
in the large N approximation are in fact also simpler to analyse and
understand.In particular, the phase relationships between the various
quantities such as the zero mode, the fluctuations and the particle production
can be picked up at a glance from the plots diplaying the solutions.

By examining these plots we noticed that the frequency of oscillations
of both the fluctuations and the particle production is double the frequency
of the zero mode oscillations. You may do this simply by counting the peaks
of the oscillations in a fixed interval of time for the various quantities.
Further, notice that the peak in particle production takes place exactly
when the zero mode is at the minimum of its potential and therefore has its
maximum kinetic energy within its oscillation. Thus, particle production
 and dissipation are proportional to the magnitude of the velocity within
each oscillation of the zero mode. Within one complete oscillation of the
zero mode it passes through the lowest point in its potential twice and there
are two peaks in particle production within each complete oscillation of the
zero mode.

This clear phase relationship between the zero mode and the
particle production
seen in the large N approximation gets somewhat more complicated for the
cutoff dependent Hartree case because higher harmonics of the fundamental
frequency start playing a significant role in the time dependence.

However, in both cases if one realises that there is a fundamental time scale
in the problem related to the oscillation time scale of the zero mode, then
one can gain some further insight into the behavior of the system and draw
some further conclusions.

Thus, if one chooses to call the time-scale of one zero mode oscillation as
the microscopic time-scale
of the problem, one can average over this microscopic
time-scale to get the behavior on much larger time-scales - the macroscopic
behavior of the system. By re-examining the plots of particle production
in this light one can see that on macroscopic time-scales the
time-averaged particle production is a monotonically growing function of time
which reaches an asymptotic final value when all the modes are sufficiently
populated.

All the momentum modes are not equally populated. In fact,
energetically one expects lower momentum modes to be more populated than
higher momentum modes.
By examining the plots of N(k) versus k, one can draw the following
conclusions. One notices that there are some distinct resonances due to
favorable phase relationships at specific momenta compared to neighboring
momenta in phase space. Further, the lowest energy resonance is always
the dominant resonance because the fields are bosonic fields and energetically
it is preferable to occupy the low momentum modes.

Further, the phase space factor $ k^2 dk $ makes the contribution of the
low momentum modes to the total occupation number less significant than the
sum of occupation numbers over the higher momentum modes. This can be verified
by examining the magnitudes of the numbers for $ N(k) $ and comparing it to
the total occupation number obtained by integrating over the momenta,
$ N_{tot} $.

In all the different cases examined by us as time progresses,
the amplitude of the zero mode oscillations decreases as the
fluctuations grow
and particle production increases.The amount of particle production and
damping increase as the contribution of the cubic term in the zero mode
evolution equation becomes significant and also as the coupling is increased.

We examined the behavior of the system in both the unbroken symmetry
case as well as the broken symmetry case.In the O(N) model in the large
N approximation one can see a very striking difference between the
unbroken symmetry case and the broken symmetry case. Whereas in the unbroken
symmetry case there is {\it some} diminishing of the amplitude of the zero
mode oscillations due to dissipation, the zero mode oscillations never
completely die out. However, for the broken symmetry case there is a very
rapid damping of the oscillations of the zero mode about its central value.
The amplitude of oscillations about this central value goes almost completely
to zero very quickly.

Whereas for moderate couplings the zero mode settles asymptotically to a
non-zero final value in the broken symmetry case there is a further interesting
twist in the extremely weak coupling case. Thus for instance for $ g \sim
10^{-7} $ what happens is that fluctuations grow by such a large amount
in the time that the zero mode is still above the spinodal point that
after making a brief excursion to non-zero values the zero mode returns to
a zero value. This happens because it is driven there by the fluctuations term
in the evolution equation.

Thus, we see in all cases
that the zero mode damps out as particle production takes place.
Further, in the broken symmetry case, the final value of the zero
mode depends on the coupling and is different from the position at the
bottom of the tree level potential.

Here then is the summary of all the results gleaned from the plots of
the solutions to the evolution equations.
The fundamental time-scale of the problem is set by the oscillation
time-scale of the zero mode when it is placed in the potential and allowed to
evolve. Within each such oscillation, particle production takes place
prefentially when the zero mode is at its lowest position in the potential
and thus has the greatest magnitude for its velocity and kinetic energy.
Further, if one keeps in mind this fundamental time-scale in the problem
and factors out all the oscillations in various quantities on this time-scale
one can follow with the eye the mean value of the quantities if one is
interested in the coarse-grained behaviour of the system. Thus, in all cases
examined, the fluctuations grow, particle production increases and the zero
mode amplitude decrease as time progresses until they reach their final
values.
The distribution of occupation number in momentum space shows the presence
of resonances. Further, the lowest energy resonance is always the dominant
resonance.

Our study also reconciles dissipation in the time evolution for the
coarse grained variable with time reversal invariance, as the evolution is
completely specified by an {\it infinite set} of ordinary second order
differential
equations in time with proper boundary conditions.

Our formalism and techniques are sufficiently powerful to give a great deal of
insight into the particulars of the dissipation process. In particular, we can
see that the damping of the field expectation value ends as the particle
production ends. This
shows that our interpretation of the damping as being due to particle
production
is accurate.

It is useful to compare what we have done here with other work on this issue.
We have already mentioned the work of Calzetta, Hu\cite{calzettahu} and
Paz\cite{paz}. These authors use the closed
time path formalism to arrive at the effective equations of motion for the
expectation value of the field.
Then they solve the {\it perturbative} equations and find dissipative evolution
at short times. In particular Paz finds the kernel that we have found for the
effective equations of
motion of the expectation value in the perturbative and Hartree case. However,
his
perturbative solution is not consistent.

We remedy this situation by studing the non-perturbative Hartree equations,
which
must necesarily be solved numerically as we do.

There has been other, previous, work on the reheating problem, most notably by
Abbott, Farhi and Wise\cite{abbottwise}, Ringwald\cite{ringwald} and
Morikawa and Sasaki\cite{morikawa}. In all of these works, the standard
effective action is used, so that the expectation value is of the ``in-out''
type and hence the equations are non-causal and contain imaginary parts.
In essence, they ``find'' dissipational behavior by adding an imaginary part
to the frequency that appears in the mode equations. We see from our work
(as well as that of Calzetta, Hu and Paz) that this is not necessary;
dissipation can occur even when
the system is evolving unitarily. This comment deserves a definition of what
we call dissipation here: it is the energy transfer from the expectation
value of the $q=0$ mode of the scalar field to the quantum fluctuations
($q \neq 0$) resulting in damped evolution for the $q=0$ mode.

To what extent are we truly treating the reheating problem of inflationary
models? As stated in the introduction, reheating typically entails the decay of
the inflaton into lighter particles during its oscillations.
What we do here is
understand how the quantum fluctuations and the ensuing particle production
influence the dynamics of the evolution  of the expectation value of the
field. Thus technically speaking, this is not the reheating problem.
However, we
{\em are} able to understand where dissipation comes from in a field theory,
and are able to give a quantitative description of the damping process for
the expectation value of the field.

The techniques we develop here are easily adapted to
the case where the inflaton couples to fermions and
also to other scalars, a case
that we have also studied\cite{us}. In this connection,
during the course of this work, two related pieces of work on the reheating
problem have appeared\cite{lindebranden}. They both look at the effect of
particle production from the oscillations of the inflaton field due to
parametric amplification. What they do {\em not} do is to account for the
back reaction of the produced particles on the evolution of the expectation
value
of the inflaton.
 As we have learned with our study, this back reaction
 will eventually shut off the particle production, so that
these authors may have overestimated the amount of particle production.

So far our in our analysis of dissipation we have been working
in Minkowski space. We must extend our analysis to an expanding
universe. This is perhaps the most interestinng and promising
future direction in the study of the dissipation of vacuum energy
following a phase transition. Some of the machinery for this extension
has already been developed\cite{frw}. We now need to do a detailed
numerical analysis of the process of dissipation via particle production
in curved spacetime.

Through the detailed applications presented in this chapter and
in earlier works we hope we have taken another step in making
finite temperature field theory in both the imaginary time and
real time formalisms a part of the toolkit of the working
cosmologist.
Applications of the real time formalism to the study of other
time dependent phenomena such as thermal activation\cite{thermact}
and domain formation and growth\cite{boysinlee} are also available
for the reader who wihses to see more detailed examples.
In this connection, extension of these studies to curved space-time,
in particular, for domain formation and growth is again a useful and
interesting direction for future investigations.

\newpage


\newpage

{\bf Figure Captions}

Note: In all figures that involve time recall that $\tau$ is the dimensionless
time defined in this chapter as $\tau = m_R t$

Fig.1: Damping of the oscillations of the field about the minimum of
its potential

Fig.2: First order quantum correction for discrete symmetry case $\eta_1(\tau)$
for $\eta_1(0)=0 \; ; \; \dot{\eta}_1(0)=0
\; ; \; \eta_{cl}(0)=1 \; ; \; \dot{\eta}_{cl}(0)=0$. The cutoff is
$\Lambda/m_R =100$.

Fig.3: Tree level potential in the unbroken symmetry case.
(For purposes of illustration we have shown the potential:
$V(\phi) = \frac{\phi^2}{2}+\frac{\phi^4}{4!}$ )

Fig. 4: Tree level potential in the broken symmetry case.
(For purposes of illustration we have shown the potential:
$V(\phi) = -\frac{\phi^2}{2}+\frac{\phi^4}{4!}$ )

Fig. 5.a: $\eta(\tau)$ vs $\tau$  in the Hartree approximation, unbroken
symmetry
case. $g=0.1 \; ; \; \eta(0)=1 \; ; \; \Lambda / M_R=100$.

Fig. 5.b: $\Sigma(\tau)$ vs $\tau$ for the same case as in fig. (5.a).

Fig. 5.c: $N(\tau)$ vs $\tau$ for the same case as in fig. (5.a).

Fig. 6.a: $\eta(\tau)$ vs $\tau$  in the Hartree approximation, unbroken
symmetry
case. $g=0.1 \; ; \; \eta(0)=4 \; ; \; \Lambda / M_R=100$.

Fig. 6.b: $\Sigma(\tau)$ vs $\tau$ for the same case as in fig. (6.a).

Fig. 6.c: $N(\tau)$ vs $\tau$ for the same case as in fig. (6.a).

Fig. 7.a: $\eta(\tau)$ vs $\tau$  in the Hartree approximation, unbroken
symmetry
case. $g=0.1 \; ; \; \eta(0)=5 \; ; \; \Lambda / M_R=100$.

Fig. 7.b: $\Sigma(\tau)$ vs $\tau$ for the same case as in fig. (7.a).

Fig. 7.c: $N(\tau)$ vs $\tau$ for the same case as in fig. (7.a).

Fig. 8.a: $\eta(\tau)$ vs $\tau$  in the Hartree approximation, unbroken
symmetry
case. $g=0.05 \; ; \; \eta(0)=5 \; ; \; \Lambda / M_R=100$.

Fig. 8.b: $\Sigma(\tau)$ vs $\tau$ for the same case as in fig. (8.a).

Fig. 8.c: $N(\tau)$ vs $\tau$ for the same case as in fig. (8.a).

Fig. 9.a: $\eta(\tau)$ vs $\tau$  in the Hartree approximation, broken symmetry
case. $g=10^{-5} \; ; \; \eta(0)=10^{-5} \; ; \; \Lambda / \mu_R=100$.

Fig. 9.b: $\Sigma(\tau)$ vs $\tau$ for the same case as in fig. (9.a).

Fig. 9.c: $N(\tau)$ vs $\tau$ for the same case as in fig. (9.a).

Fig. 10.a: $\eta(\tau)$ vs $\tau$  in the large N approximation in the O(N)
model, unbroken symmetry
case. $g=0.1 \; ; \; \eta(0)=1 \; ; \; \Lambda / M_R=100$.

Fig. 10.b: $\Sigma(\tau)$ vs $\tau$ for the same case as in fig. (10.a).

Fig. 10.c: $N(\tau)$ vs $\tau$ for the same case as in fig. (10.a).

Fig. 11.a: $\eta(\tau)$ vs $\tau$  in the large N approximation in the O(N)
model, unbroken symmetry
case. $g=0.1 \; ; \; \eta(0)=2 \; ; \; \Lambda / M_R=100$.

Fig. 11.b: $\Sigma(\tau)$ vs $\tau$ for the same case as in fig. (11.a).

Fig. 11.c: $N(\tau)$ vs $\tau$ for the same case as in fig. (11.a).

Fig. 12.a: $\eta(\tau)$ vs $\tau$  in the large N approximation in the O(N)
model, unbroken symmetry
case. $g=0.3 \; ; \; \eta(0)=1 \; ; \; \Lambda / M_R=100$.

Fig. 12.b: $\Sigma(\tau)$ vs $\tau$ for the same case as in fig. (12.a).

Fig. 12.c: $N(\tau)$ vs $\tau$ for the same case as in fig. (12.a).

Fig. 13.a: $\eta(\tau)$ vs $\tau$  in the large N approximation in the O(N)
model, broken symmetry
case. $g=0.1 \; ; \; \eta(0)=0.5 \; ; \; \Lambda / \mu_R=100$.

Fig. 13.b: $\Sigma(\tau)$ vs $\tau$ for the same case as in fig. (13.a).

Fig. 13.c: $N(\tau)$ vs $\tau$ for the same case as in fig. (13.a).

Fig. 13.d: Number of particles in (dimensionless) wavevector $q$, $N_q(\tau)$
at
$\tau=200$.

Fig. 14.a: $\eta(\tau)$ vs $\tau$  in the large N approximation in the O(N)
model, broken symmetry
case. $g=10^{-7} \; ; \; \eta(0)=10^{-5} \; ; \; \Lambda / \mu_R=100$.

Fig. 14.b: $\Sigma(\tau)$ vs $\tau$ for the same case as in fig. (14.a).

Fig. 14.c: $N(\tau)$ vs $\tau$ for the same case as in fig. (14.a).

Fig. 15.a: $\eta(\tau)$ vs $\tau$  in the large N approximation in the O(N)
model, broken symmetry
case. $g=10^{-12} \; ; \; \eta(0)=10^{-5} \; ; \; \Lambda / \mu_R=100$.

Fig. 15.b: $\Sigma(\tau)$ vs $\tau$ for the same case as in fig. (15.a).

Fig. 15.c: $N(\tau)$ vs $\tau$ for the same case as in fig. (15.a).


\end{document}